\documentclass[pdflatex,sn-mathphys-num]{sn-jnl}% Math and Physical Sciences Numbered Reference Style
\usepackage{sansmath}
\usepackage{graphicx}%
\usepackage{multirow}%
\usepackage{amsmath,amssymb,amsfonts}%
\usepackage{amsthm}%
\usepackage{mathrsfs}%
\usepackage[title]{appendix}%
\usepackage{xcolor}%
\usepackage{textcomp}%
\usepackage{manyfoot}%
\usepackage{booktabs}%
\usepackage{algorithm}%
\usepackage{algorithmicx}%
\usepackage{algpseudocode}%
\usepackage{listings}%
\usepackage{natbib}
\usepackage{longtable}
\usepackage{tcolorbox}
\usepackage{lineno}
\usepackage[table]{xcolor}
\usepackage{multicol}
\usepackage{comment}
\usepackage{amsmath} % For \dfrac
\usepackage{booktabs} % For professional-looking tables (\toprule, \midrule, \bottomrule)
\usepackage{multirow} % To span rows

\usepackage{cleveref}

\newcounter{extfig}
\newcounter{exttable}

\crefname{extfig}{Extended Data Fig.}{Extended Data Figs.}
\Crefname{extfig}{Extended Data Fig.}{Extended Data Figs.}

\crefname{exttable}{Extended Data Table}{Extended Data Tables}
\Crefname{exttable}{Extended Data Table}{Extended Data Tables}

\newcommand{\extfiglegend}[3]{%
    \refstepcounter{extfig}% Increment the counter
    \label{#1}% Register the label for referencing
    \noindent\textbf{Extended Data Fig. \theextfig\ | #2.} #3% Print Title and Text
    \vspace{1em}%
}

\theoremstyle{thmstyleone}%
\theoremstyle{thmstyletwo}%

\theoremstyle{thmstylethree}%

\begin{document}
% \linenumbers

\title[Article Title]{The Rise and Fall of ENSO in a Warming World: Insights from a Lag-Linear Model}

% Title options:

\author*[1,2,3]{\fnm{P.J.} \sur{Tuckman}}\email{pjt5@stanford.edu}

\author*[1,2]{\fnm{Da} \sur{Yang}}\email{dayang@stanford.edu}

\affil[1]{\orgdiv{Department of Geophysical Sciences}, \orgname{University of Chicago}, \orgaddress{\street{5734 S Ellis Ave}, \city{Chicago}, \state{IL}, \postcode{60637}, \country{United States of America}}}

\affil[2]{\orgdiv{Department of Geophysics}, \orgname{Stanford University}, \orgaddress{\street{397 Panama Mall}, \city{Stanford}, \state{CA}, \postcode{94305},  \country{United States of America}}}

\affil[3]{\orgname{University Corporation for Atmospheric Research}, \orgaddress{\street{3090 Center Green Dr}, \city{Boulder}, \state{CO}, \postcode{80301},  \country{United States of America}}}

% \affil[2]{\orgdiv{Department of Earth, Atmospheric, and Planetary Sciences}, \orgname{Massachusetts Institute of Technology}, \orgaddress{\street{21 Ames Street}, \city{Cambridge}, \postcode{02139}, \state{MA}, \country{United States of America}}}


\abstract{The El Ni\~no-Southern Oscillation (ENSO) is a fluctuation in sea surface temperature and pressure across the equatorial Pacific Ocean with a period of 2-7 years. As the largest mode of interannual variability on Earth, ENSO shapes global weather and climate patterns ranging from monsoons in southern Asia to hurricanes in the Atlantic and droughts in South America. Predicting and understanding ENSO's response to greenhouse warming is essential for mitigating the impacts of climate change, yet model ensemble projections are expensive to generate across emission scenarios and remain incompletely understood. Here, we use a hierarchy of models to explain the transient rise and subsequent fall of ENSO strength under greenhouse warming, then develop an efficient and accurate method for predicting ENSO variability in any emissions scenario. Beginning with an East Pacific energy budget, we quantitatively show how enhanced upper-ocean stratification strengthens ENSO and how a slowing Walker circulation and stronger surface flux damping eventually weaken it. This leads to a linear model that predicts the evolution of ENSO variability from only East Pacific temperature and stratification. We further show that subsurface warming, and therefore stratification, is connected to surface warming with a lag, enabling us to create a lag-linear model that explains $\sim$90\% of simulated changes in ENSO variability from only global mean surface temperature and its history. Once calibrated, this efficient predictor can project ENSO strength without running a full climate model, and allows an analytic solution for the timing and magnitude of peak ENSO variability. We find that the ratio of an ocean subsurface adjustment timescale to the warming timescale strongly alters peak ENSO amplitude, meaning that faster emissions lead to larger ENSO variability even with identical total emissions.}




\maketitle

%\section*{Main}\label{sec:Introduction}

%El Ni\~no events, months-long warm anomalies in the equatorial East Pacific,

\begin{comment}
\begin{tcolorbox}[colback=yellow!30, colframe=white, arc=0pt, boxrule=0pt]
This entire section or paragraph is highlighted. 

You can include multiple paragraphs, math equations, and lists inside this box, and it will perfectly format as a highlighted block of text.
\end{tcolorbox}

\end{comment}

\begin{figure}[h]
\centering
\includegraphics[width=0.9\textwidth]{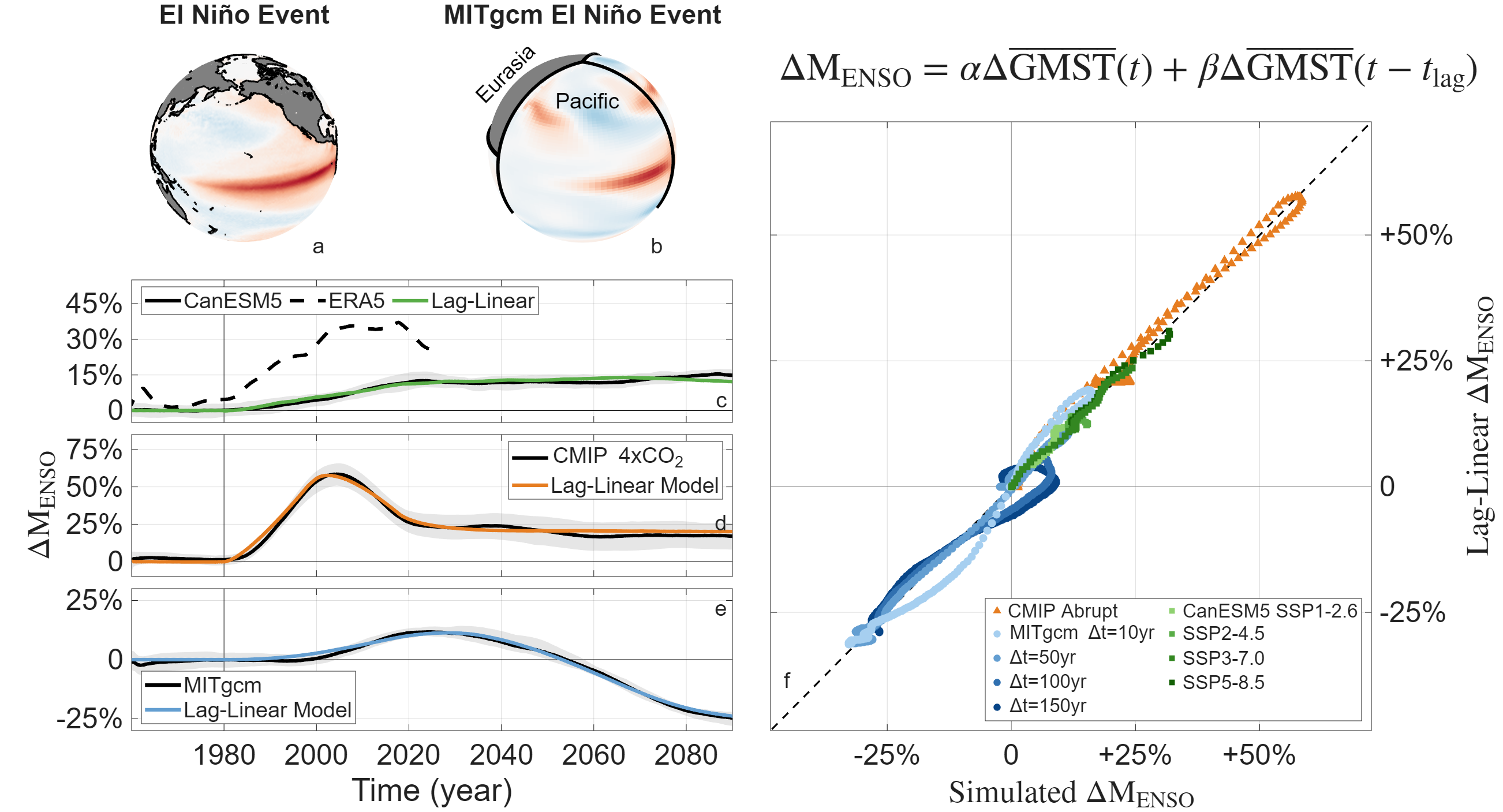}
\caption{A lag-linear model capturing the rise and fall of ENSO variability. Panels a and b show El Ni\~no events from ERA5 (a) and the MITgcm idealized simulation (b). Panels c-e show $\Delta \mathrm{M}_{\mathrm{ENSO}}$, i.e., the normalized change in ENSO strength smoothed with a 20 year causal moving mean, and predictions based on the lag-linear model from ERA5 (c, dashed line), CanESM5 under scenario SSP3-7.0 (c, solid line), CMIP models under an abrupt quadrupling of CO$_2$ (defined to occur in 1980, d), and idealized MITgcm simulations (e). The top right of the figure shows the equation for the lag-linear model, where $\Delta \overline{\mathrm{GMST}}$ is the change in global mean SST and $\alpha$, $\beta$, and $t_{\mathrm{lag}}$ are constant parameters. Panel f shows the relationship between the lag-linear prediction and the simulated changes in ENSO strength in all simulations studied (a CMIP ensemble, four MITgcm scenarios with strengthening greenhouse warming over differing warming timescales $\Delta t$, and four CanESM5 scenarios). The lag-linear parameters are allowed to vary across models, but are held constant across warming scenarios. The same quantities without the 20-year smoothing are shown in Extended Data Fig. \ref{Fig:ED_UnsmoothedENSO}. Uncertainty bands in each panel are calculated as the 2.5\%-97.5\% of an N=100 bootstrap resampling of the data averaged in that panel.}\label{Fig:IntroFigure}
\end{figure}

% impacting marine resources \citep{lehodey_climate_2006}, altering disease spread dynamics \citep{hales_nino_1999}, 
Every 2-7 years, the equatorial East Pacific becomes anomalously warm for several months in a phenomenon known as an ``El Ni\~no event" (Fig. \ref{Fig:IntroFigure}a). These events disrupt society worldwide by lowering agricultural productivity \citep{adams_economic_1999},  causing water shortages \citep{liu_nonlinear_2023}, and modulating weather and climate events such as floods  \citep{hamlet_effects_2007}, droughts, and tropical cyclones \citep{lin_enso_2020}. The last two extreme El Ni\~no events, in 1997-1998 and 2015-2016, caused estimated economic productivity losses of up to \$4.1 and \$5.7 trillion, respectively, while their opposite, La Ni\~na events, lead to much smaller economic gains \citep{callahan_persistent_2023}. Occurrences of El Ni\~no and La Ni\~na, along with corresponding changes in sea level pressure and the Walker circulation, are collectively known as the El Ni\~no-Southern Oscillation (ENSO), and have been studied extensively over the last few decades \citep[e.g.,][]{bjerknes_atmospheric_1969,wyrtki_water_1985,battisti_understanding_1995,mcphaden_enso_2006,timmermann_ninosouthern_2018}. As the Earth warms during the 21st century, ENSO is expected to cause tens of trillions of dollars of lost economic output \citep{liu_nonlinear_2023,callahan_persistent_2023}, yet the response of ENSO to different greenhouse gas emission scenarios remains expensive to predict and difficult to interpret \citep{fedorov_is_2000,latif_ninosouthern_2009,intergovernmental_panel_on_climate_change_ipcc_climate_2023,lee_future_2021}.

ENSO variability is measured as the temporal standard deviation of equatorial East Pacific upper-ocean temperatures \citep{guilyardi_understanding_2009}. We represent changes in ENSO strength using:
\begin{equation}
    \Delta \mathrm{M}_{\mathrm{ENSO}}(t) = \frac{\text{East Pacific Temperature Standard Deviation}}{\text{Control East Pacific Temperature Standard Deviation}}-1
\end{equation}
and smooth the result with a 20-year causal moving mean to remove unrelated internal variability. The ``Control East Pacific Temperature Standard Deviation" represents ENSO variability in a reference climate (e.g., 1960-1980) and is used so that $\Delta \mathrm{M}_{\mathrm{ENSO}}$ is zero before warming and positive when ENSO events are more extreme.

Climate model ensembles indicate that there will be a transient rise and long-term fall in ENSO variability under greenhouse warming \citep{kim_response_2014,callahan_robust_2021,geng_decreased_2024,maher_future_2023}. In observations, ENSO events strengthened during the final decades of the 20th century (Fig. \ref{Fig:IntroFigure}c), likely in part due to anthropogenic emissions \citep[][]{hersbach_era5_2020,cai_changing_2021,gan_greenhouse_2023}. Most models predict that this enhanced ENSO variability will continue in the 21st century, possibly due to greenhouse warming increasing upper-ocean stratification \citep[Fig. \ref{Fig:IntroFigure}c,][]{cai_increased_2018,cai_changing_2021,heede_towards_2023}, though model predictions of ENSO still have substantial spread \citep{cai_increased_2015,wang_continued_2017,cai_increased_2018}. This rise is temporary and ENSO variability is expected to eventually decrease with warming \citep{kim_response_2014}, most clearly seen in simulations after an abrupt CO$_2$ increase (Fig. \ref{Fig:IntroFigure}d). Idealized models under warming (Fig. 
\ref{Fig:IntroFigure}b and e) also show a transient peak of ENSO variability followed by a lasting decline, and simulations run past 2100 \citep{peng_collapsed_2024,geng_decreased_2024,callahan_robust_2021} or to equilibrium \citep{tuckman_understanding_2025} often show weaker ENSO events in warmer climates.

Here, we use idealized and comprehensive climate models and an East Pacific energy budget to derive a lag-linear model for ENSO strength in terms of global mean SST (GMST, equation in Fig. \ref{Fig:IntroFigure}). This model accurately reproduces the rise and fall of ENSO variability, capturing around $90\%$ of the signal across models and warming scenarios (Fig. \ref{Fig:IntroFigure} time series and panel f). We use this simple predictor to gain insight into the physical process controlling the future of ENSO events, finding that peak ENSO variability depends on the warming magnitude, the ratio of ENSO's sensitivities to surface and subsurface changes, and the ratio of an ocean subsurface adjustment timescale to the warming timescale. 

\clearpage

\section*{Fast El Ni\~no and Slow La Ni\~na Responses to Warming}\label{sec:Results}

We begin by showing that ENSO variability rises under warming due to differing changes in El Ni\~no and La Ni\~na events; greenhouse emissions cause an immediate rise in El Ni\~no event temperatures, corresponding to ocean surface warming, while La Ni\~na event temperatures, which are influenced by the subsurface, remain cool for several decades after warming begins (Extended Data Fig. \ref{Fig:ED_ElNinoVsLaNina}). To demonstrate this, we use three sets of simulations: 1. an ensemble of CMIP models after an abrupt quadrupling of CO$_2$ \citep{eyring_overview_2016}; 2. a Community Earth System Model Large Ensemble undergoing the SSP3-7.0 scenario \citep[CESM2,][]{rodgers_ubiquity_2021}; and 3. idealized atmosphere-ocean MITgcm simulations in which atmospheric longwave absorption is increased over 10, 50, 100, or 150 years \citep[continental configuration shown in Fig. \ref{Fig:IntroFigure} and Extended Data Fig. \ref{Fig:ED_ElNinoVsLaNina},][]{marshall_finite-volume_1997,tuckman_enso_2025}. 

In the abrupt warming simulations, 20-year maximum East Pacific temperatures (i.e., the temperature during strong El Ni\~no events) rise immediately after the CO$_2$ increase, while 20-year minimum East Pacific temperatures (corresponding to strong La Ni\~na events) remain nearly constant for two decades (Extended Data Fig. \ref{Fig:ED_ElNinoVsLaNina}a.i, dark red vs. dark blue line). Similarly, the CESM2 and MITgcm simulations show that minima temperature warming occurs two to three decades after maxima temperature warming (Extended Data Fig. \ref{Fig:ED_ElNinoVsLaNina}a). 

Temperature maxima increase before temperature minima because ENSO alters equatorial East Pacific upwelling. Temperatures during El Ni\~no events, when upwelling is suppressed and zonal asymmetries across the tropical Pacific are small, warm at the same rate as tropical mean near-surface water (Extended Data Fig. \ref{Fig:ED_ElNinoVsLaNina}b, red lines). La Ni\~na events, on the other hand, enhance upwelling, so their rise in temperature is slowed by cooler subsurface water entering the mixed layer (Extended Data Fig. \ref{Fig:ED_ElNinoVsLaNina}b, blue lines). In other words, as atmospheric temperatures increase, the near-surface ocean warms before the subsurface, leading to increased El Ni\~no temperatures and temporarily enhanced ENSO variability. The importance of contrasting near- and subsurface behavior suggests that the rise and fall of ENSO strength shown in Fig. \ref{Fig:IntroFigure} depend strongly on stratification \citep[consistent with previous work:][]{cai_increased_2018,cai_changing_2021}; we now study this quantitatively with an equatorial East Pacific energy budget.

\clearpage

\section*{Energy Balance of the Equatorial East Pacific}
\label{subsec:EnergyBudget}

\begin{figure}[h]
\centering
\includegraphics[width=0.9\textwidth]{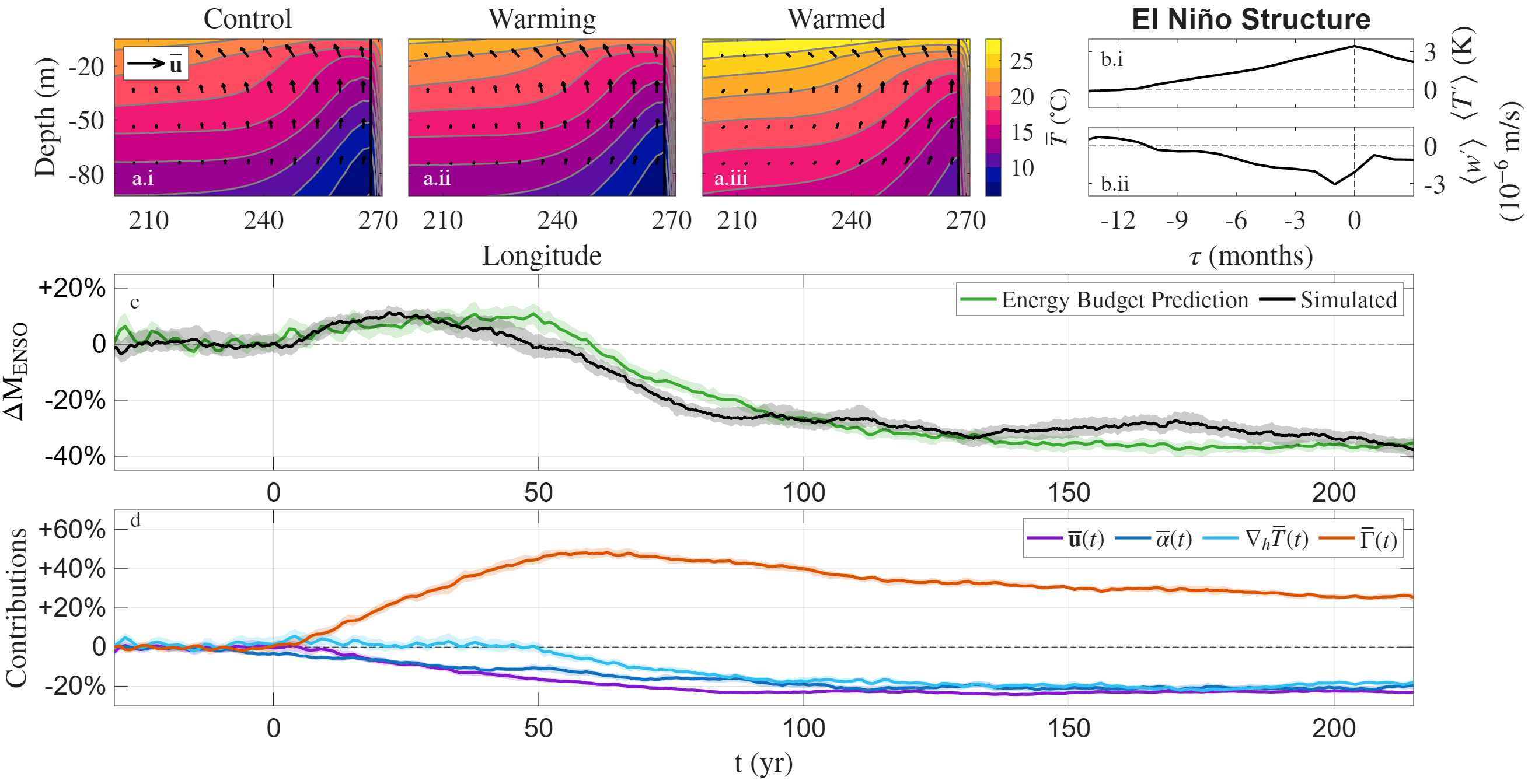}
\caption{Using energy balance to study how the magnitude of El Ni\~no events changes under warming in the MITgcm $\Delta t=$ 50 years simulations. Panel a shows the mean temperature (color) and currents (arrows) in the pre-warming climate (a.i), the transient warming state (a.ii), and the (mostly) equilibrated warmed state (a.iii). Panel b shows the volume averaged anomalous temperature (b.i) and vertical velocity (b.ii) from a composite El Ni\~no structure in the reference climate. Panel c shows the predicted $\Delta \mathrm{M}_{\mathrm{ENSO}}$ from energy balance and compares it to the simulated results. Panel d shows how specific terms affect the predicted ENSO magnitude. Uncertainty bands are calculated by applying an N=100 bootstrap resampling to the input bar variables then using the 2.5\%-97.5\% values to predict ENSO peaks, and the central line displayed is the bootstrap median.}\label{Fig:EnergyBudget}
\end{figure}

We use an energy budget of the equatorial East Pacific to reproduce simulated changes of ENSO variability in the MITgcm, showing that ENSO's transient rise is associated with enhanced upper-ocean stratification and that its eventual decline is associated with decreasing stratification, a weakening Walker circulation, and strengthening surface flux damping (Fig. \ref{Fig:EnergyBudget}). The anomalous temperature of the mixed layer is given by:

\begin{equation}
\label{eq:FirstBudget}
    \frac{d}{d t} \langle T' \rangle = -\langle \bar{\mathbf{u}} \cdot \nabla T' \rangle - \langle \mathbf{u}_h' \cdot \nabla_h \bar{T} \rangle - \langle w' \bar{\Gamma} \rangle -\bar{\alpha}_{\mathrm{SF}} \langle T' \rangle_{\text{2D,surf}}
\end{equation}
where $\langle \cdot \rangle$ indicates a volume average over the equatorial East Pacific ($\sim210-270^\circ$E, within $\sim5$ degrees of the equator, top 80 m), $T$ is temperature, $\mathbf{u}$ is the three-dimensional (3D) current, $\nabla$ is the 3D gradient, the $h$ subscript represents horizontal quantities, $w$ is the vertical current, $\Gamma$ is the stratification ($\partial T/\partial z$), bar variables (e.g., $\overline{T}$) represent the average climate through 20-year moving means, and prime variables (e.g., $T'$) represent ENSO anomalies as deviations from this temporal mean with the seasonal cycle removed. The association of $\langle T' \rangle$ with ENSO is supported by a spectral analysis showing that most anomalous East Pacific temperature variability occurs in the 2-7 year ENSO band. The final term of Eq. \ref{eq:FirstBudget} treats anomalous surface fluxes and longwave radiation (i.e., thermodynamic damping) as the product of a slowly varying coefficient $\bar{\alpha}_{\mathrm{SF}}$ and the anomalous regional SST. For clarity, we here omit terms that do not significantly affect ENSO strength such as mixing and advection associated with the seasonal cycle; they are included in the calculations (details in the methods section). Additionally, we treat the mixed layer as a fixed volume, though the mean thermocline depth may change over time.

%We use Eq. \ref{eq:FirstBudget} to create an expression for ENSO variability in a given climate. Assuming a composite across several El Ni\~no events occurring in that climate and defining a time variable $\tau$ such that the events peak when $\tau=0$, 

We turn Eq. \ref{eq:FirstBudget} into an expression for the magnitude of ENSO peaks in a given climate by dividing both sides by $\langle T' \rangle$, integrating from the initiation of an El Ni\~no event to its peak, and normalizing by a reference budget (details in methods section):
\begin{equation}
\label{Eq:BudgetMainText}
    \Delta \mathrm{M}_{\mathrm{ENSO}} (t) = \exp\!\left[\int_{\tau_0}^0 \Delta\sigma d\tau\right]-1,
\end{equation}
where 
\begin{equation}
\label{Eq:BudgetMainTextSigma}
    \Delta \sigma  \equiv -\frac{\langle \Delta \mathbf{\bar{u}}\!\cdot\!\nabla T' \rangle}{\langle T' \rangle}-\frac{\langle \mathbf{u'}_h \cdot \Delta \nabla_h \bar{T} \rangle}{\langle T' \rangle} -\frac{\langle w' \Delta \bar{\Gamma} \rangle}{\langle T' \rangle} -\Delta \bar{\alpha}_{\mathrm{SF}}\,\frac{\langle T' \rangle_{\mathrm{2D,surf}}}{\langle T' \rangle}. 
\end{equation}
Changes in the ENSO growth rate, $\Delta \sigma$, represent the instability of the tropical Pacific to El Ni\~no perturbations \citep[similar to the commonly used Bjerknes index,][discussed below]{jin_coupled-stability_2006}. In these expressions, $\Delta$ represents a change from the reference value, $\tau_0$ marks the initiation of an ENSO event defined by a fixed small value of $\langle T' \rangle$, and $\tau=0$ marks the event's peak. By setting this expression equal to $\Delta \mathrm{M}_{\mathrm{ENSO}}$, we assume that $\langle T' \rangle$ peak values are proportional to East Pacific standard deviation; this is supported by simulation results (Fig. S1). To evaluate changes in ENSO growth rate, the climatological variables in Eq. \ref{Eq:BudgetMainTextSigma} (e.g., $\bar{\Gamma}$) are allowed to vary with warming while the structure of an El Ni\~no event (represented by, e.g., $w'(\mathbf{x},\tau)/\langle T' \rangle$ and $\tau_0$) is held fixed \citep[as in the Bjerknes index,][]{jin_coupled-stability_2006}. This allows us to understand how mean climate variables alter ENSO strength by placing the same El Ni\~no event in a changing climate (details of this calculation and demonstration of constant ENSO structure are shown in methods section).

%Eq. \ref{Eq:BudgetMainText} shows that ENSO variability depends on mean climate variables such as $\mathbf{\bar{u}}$, $\bar{T}$, and $\bar{\Gamma}$; we now examine how those variables evolve in the MITgcm simulations. 

Under climate change, upper-ocean currents weaken while stratification first increases then decreases (Fig. \ref{Fig:EnergyBudget}). In the reference climate (Fig. \ref{Fig:EnergyBudget}a.i) water moves west near the surface, leading to upwelling and cooler water in the East Pacific. Once greenhouse warming begins (a.ii), surface temperatures increase significantly (by $\sim 4$K) while the subsurface has not yet experienced much warming ($\sim$1-2 K), enhancing stratification (Fig. S2). Over time, the subsurface ocean warms as well, and the stratification settles to an equilibrium value (a.iii). In this warmer climate, mean currents have decreased (especially near the surface), associated with a slower Walker circulation leading to weaker trade winds \citep[Fig. S3;][]{meehl_nino-like_1996,vecchi_global_2007,wills_local_2017,tuckman_understanding_2025}. 

The structure of ENSO events, meanwhile, is calculated from a composite across 25 El Ni\~no events in the reference climate, identified as the largest maxima of $\langle T' \rangle$ separated by at least two years. In this composite, volume averaged anomalous temperature ($\langle T' \rangle$, panel b.i) grows from zero to about 3 K over the $\sim11$ months before a peak, and the anomalous vertical velocity ($\langle w' \rangle$, b.ii) becomes strongly negative.

Combining the evolution of mean variables and a fixed ENSO structure, our energy budget predicts a transient rise and long-term fall in ENSO strength (Fig. \ref{Fig:EnergyBudget}c), matching the simulations in showing that $\Delta \mathrm{M}_{\mathrm{ENSO}}$ increases to a maximum of about +10\% between years 0 and 50, then slowly decreases to about $-40$\%. 

Eq. \ref{Eq:BudgetMainText} is similar to the Bjerknes index \citep{jin_coupled-stability_2006,jin_simple_2020} in that it uses an anomalous energy budget to predict ENSO variability by assuming a fixed El Ni\~no structure while allowing mean variables to evolve as the climate changes. This method is different from the Bjerknes index in that the quantities are volume averaged after multiplication (i.e., we use $\langle u' \nabla_x\bar{T}\rangle $ rather than $\langle u' \rangle \langle \nabla_x \bar{T} \rangle$) and we allow El Ni\~no structure (e.g., $\textbf{u}' /\langle T' \rangle$) to vary across phases of the event ($\tau$). This leads to a more accurate but complex predictor of ENSO strength in an evolving climate.

Using Eq. \ref{Eq:BudgetMainText}, we can attribute the evolution of El Ni\~no strength to individual variables by changing only that variable over time. The Walker circulation weakens with warming \citep[Fig. S3;][]{meehl_nino-like_1996,vecchi_global_2007,wills_local_2017,tuckman_understanding_2025}, leading to the mean velocity term ($\mathbf{\overline{u}}$) and the mean horizontal temperature gradient term ($\nabla_h \overline{T}$) contributing less to ENSO growth (Fig. \ref{Fig:EnergyBudget}d). The change in the mean velocity term is mostly controlled by a weakening of $\bar{w}$ (i.e., changes in the thermocline feedback), but also has a contribution from a weakening of horizontal currents (less dynamical damping, not shown). The surface flux term ($\bar{\alpha}_{\mathrm{SF}}$, thermodynamic damping) also causes a weakening of ENSO variability over time due to the exponential relationship between saturation humidity and temperature. As the saturation specific humidity of a parcel of air increases exponentially with warming, the same relative humidity corresponds to a larger saturation deficit, and therefore stronger evaporation, in a warmer climate (see Supplementary Text 1 for details). While Fig. \ref{Fig:EnergyBudget} shows quantities from the $\Delta t=50$ yr simulations only, the sign and relative amplitude of each contribution are similar across warming timescales (Fig. S4).

The transient strengthening of ENSO events, meanwhile, is associated with changes in stratification (known as the vertical advective or Ekman feedback). As the surface of the ocean responds immediately to greenhouse warming while the response of the subsurface ocean is delayed (Extended Data Fig. \ref{Fig:ED_ElNinoVsLaNina}b.ii, Fig. \ref{Fig:EnergyBudget}a, and Fig. S2), upper-ocean stratification first increases over 50 years then decreases over the following centuries. These stratification changes contribute to a fast increase then slow decrease of ENSO variability, which, combined with other mean climate changes, cause a transient rise and long-term fall in ENSO strength.

\section*{Simple Predictors of ENSO Variability}
\label{subsec:LinearModel}

We now use the East Pacific energy budget to derive a linear model for ENSO variability  in terms of mean temperature and stratification (Fig. \ref{Fig:FirstLinearModel}). As the argument to the exponential in Eq. \ref{Eq:BudgetMainText} is small, we can apply a Taylor expansion to create a linear expression for $\Delta \mathrm{M}_{\mathrm{ENSO}}$ (see Supplementary Text 2 for details). Then, by multiplying the stratification term by $\langle \Delta \bar{\Gamma}\rangle/\langle \Delta\bar{\Gamma} \rangle$ and each other term by $\langle \Delta\bar{T}\rangle/\langle \Delta\bar{T} \rangle$, we find:

\begin{figure}[h]
\centering
\includegraphics[width=0.9\textwidth]{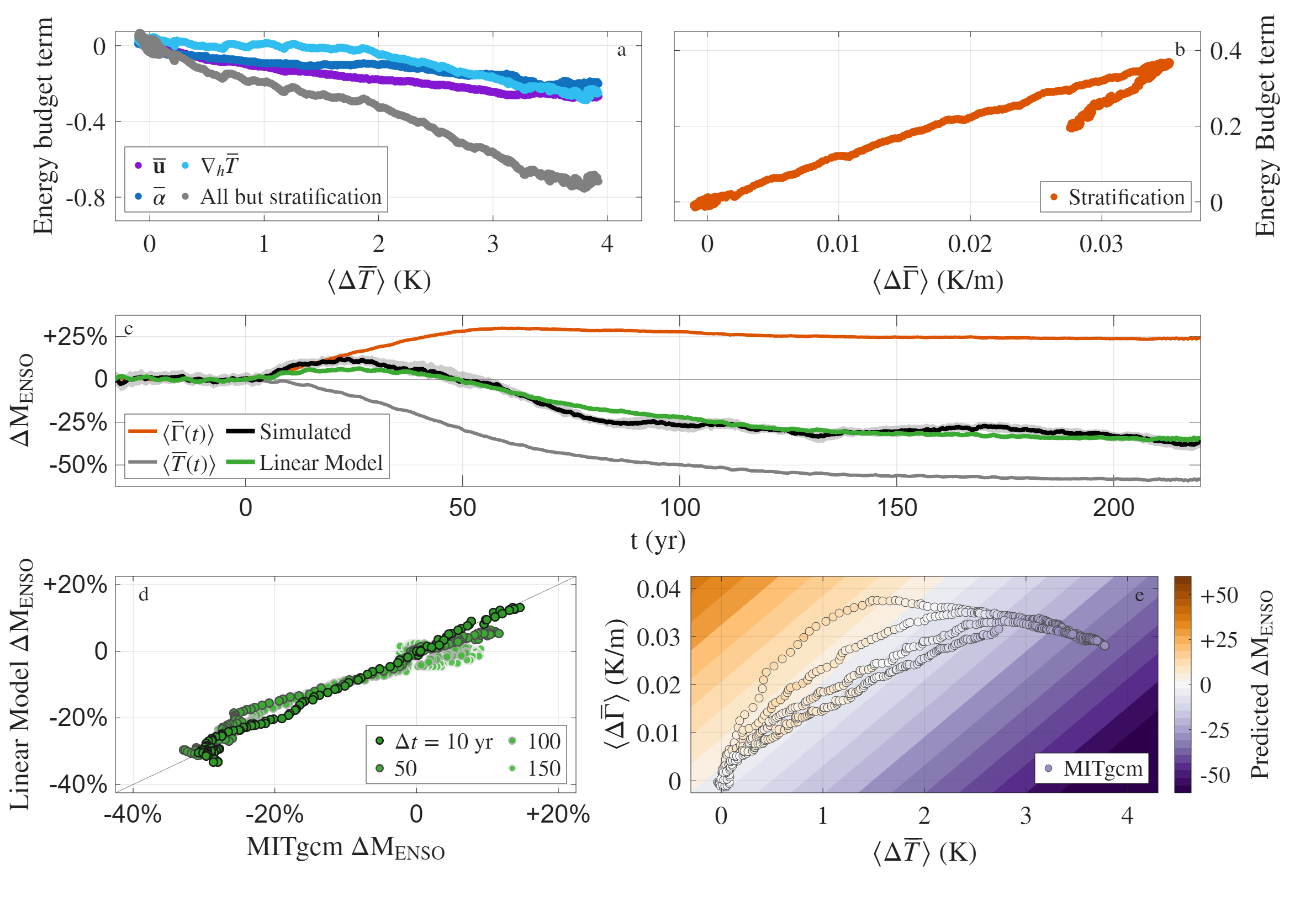}
\caption{Mean temperature and stratification control ENSO variability. Panel a shows the relationship between three of the energy budget terms and mean temperature in the MITgcm $\Delta t=50$ year simulations (Eq. \ref{Eq:APrediction}), while panel b shows the relationship between the stratification term and mean stratification (Eq. \ref{Eq:BPrediction}). Panel c shows the simulated ENSO strength from the MITgcm, the empirically fit linear model and its prediction as a function of time, as well as the predictions if only mean temperature or stratification were allowed to change. Panel d compares the simulated ENSO magnitude to that predicted by the linear model for all MITgcm simulations, where color and point outlines correspond to different warming rates. Panel e illustrates how this linear model predicts ENSO magnitude would change as a function of the two controlling variables, with ensemble mean MITgcm simulation data shown as a scatterplot.}\label{Fig:FirstLinearModel}
\end{figure}

\begin{equation}
\label{eq:StratLinearModel}
    \Delta \mathrm{M}_{\mathrm{ENSO}}(t) = A \langle \Delta \bar{T} \rangle + B \left \langle \Delta \bar{\Gamma} \right \rangle,
\end{equation}
where 

\begin{subequations} \label{Eq:CombinedLabel}
\begin{align}
    A &= - \frac{1}{\langle \Delta \bar{T} \rangle} \int_{\tau_0}^0 d\tau \left( \frac{\langle \Delta \bar{\mathbf{u}} \cdot \nabla T' \rangle}{\langle T' \rangle }+ \frac{\langle \mathbf{u}'_h\cdot \Delta \nabla_h \bar{T} \rangle}{\langle T' \rangle } +\Delta \bar{\alpha}_{\mathrm{SF}} \frac{\langle T' \rangle_{\mathrm{2D,surf}}}{\langle T' \rangle } \right) \label{Eq:APrediction} \text{   and} \\
    B &= - \frac{1}{\langle \Delta \bar{\Gamma} \rangle} \int_{\tau_0}^0 d\tau\frac{\langle w' \Delta \bar{\Gamma} \rangle}{\langle T' \rangle} .\label{Eq:BPrediction}
\end{align}
\end{subequations}
The parameters $A$ and $B$ are roughly constant in the MITgcm simulations: $A\langle \Delta \bar{T} \rangle$ and $B \left \langle \Delta \bar{\Gamma} \right \rangle$ have strong linear relationships with $\langle \Delta \bar{T} \rangle$  and $\left \langle \Delta \bar{\Gamma} \right \rangle$, respectively (demonstrated in the $\Delta t=50$ year scenario, Fig. \ref{Fig:FirstLinearModel}a and b, Extended Data Table 1). The value of $-1/A$ is around 7 K, and represents the temperature scale over which warming weakens ENSO events by a factor of $e$ (Eq. \ref{Eq:BudgetMainText}). Meanwhile, $1/B\approx0.12$ K/m is the stratification increase that would strengthen ENSO by a factor of $e$ (discussed further in Supplementary text 2). Additionally, each component corresponds to parts of the Bjerknes index (i.e., $A$ is made up of dynamical damping, zonal advective feedback, the thermocline feedback, and thermodynamic damping, while $B$ is controlled by the Ekman feedback). Overall, Eq. \ref{eq:StratLinearModel} is an energy balance based linear predictor for ENSO strength in terms of only mean temperature and mean stratification.
 
The linear model accurately predicts the evolution of simulated ENSO variability (Fig. \ref{Fig:FirstLinearModel}c). We can estimate $A$ and $B$ either by evaluating expressions \ref{Eq:APrediction} and \ref{Eq:BPrediction} (calculated as the slope of the data in Fig. \ref{Fig:FirstLinearModel}a and b) or through an empirical fit of $\Delta \mathrm{M}_{\mathrm{ENSO}}$; these two methods give similar results (Extended Data Table 1 and Fig. S5). The simplicity of the model allows for clear separation of the effects of mean temperature and stratification: warming causes a monotonic weakening of ENSO variability, while stratification changes cause a fast increase then a small decrease in ENSO strength (Fig. \ref{Fig:FirstLinearModel}c). It is important to note that the contributions of stratification changes to ENSO variability are not fully captured by a linear function of mean stratification due to changing upper-ocean temperature structure (non-linearity in Fig. \ref{Fig:FirstLinearModel}b); this leads to stratification contributing less to the decline of ENSO variability in the linear model than in the energy balance framework.

Eq. \ref{eq:StratLinearModel} can predict ENSO variability in simulations across warming scenarios with the same $A$ and $B$ coefficients (Fig. \ref{Fig:FirstLinearModel}d). In other words, as the climate warms, just two variables, mean temperature and stratification, are enough to determine ENSO variability regardless of the emissions scenario (Fig. \ref{Fig:FirstLinearModel}e). This allows for the prediction of ENSO strength across climates, even if the MITgcm is run for only one realization, simply by tracing the path of the two relevant variables.

\subsubsection*{Predicting ENSO from Surface Temperature Evolution} \label{subsubsec:LagLinear}

\begin{figure}[h]
\centering
\includegraphics[width=0.9\textwidth]{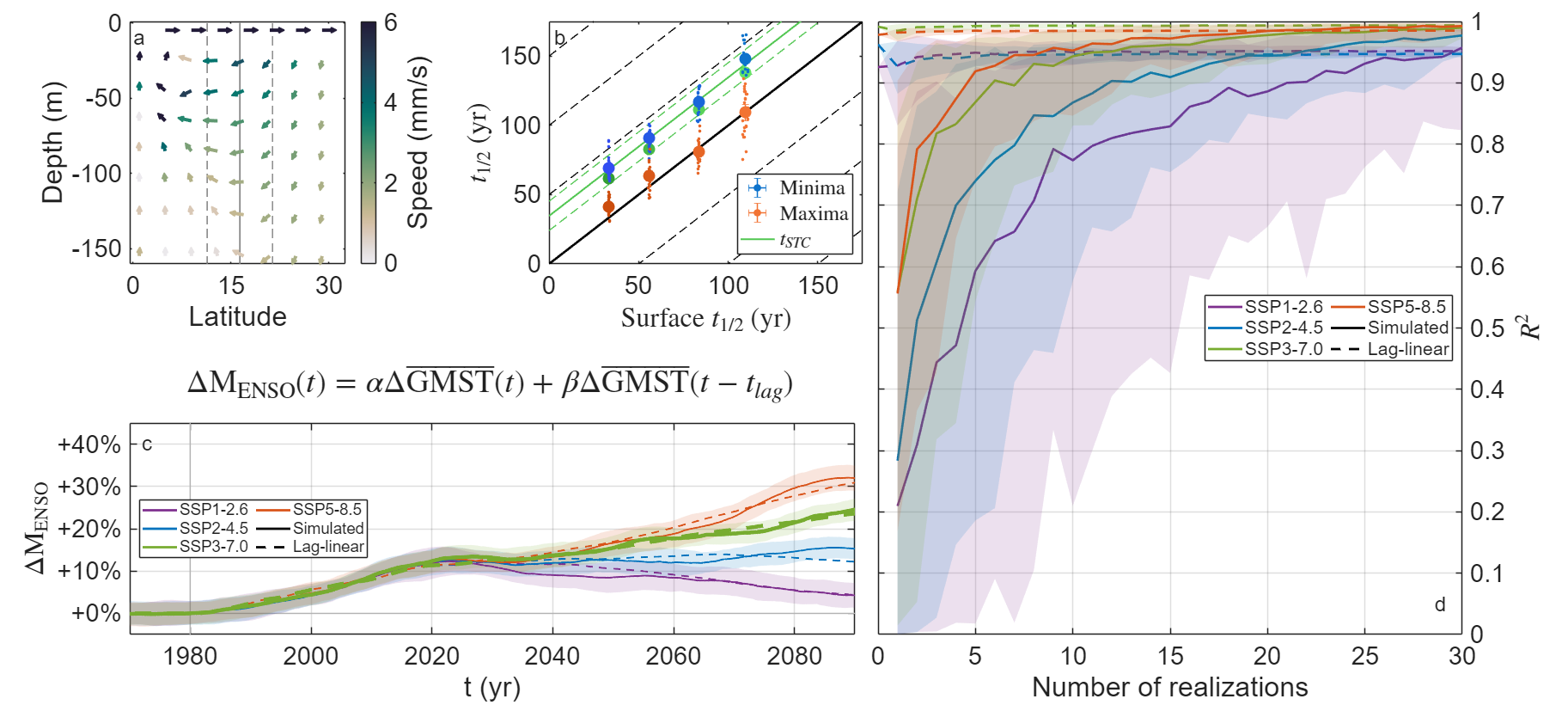}
\caption{A lag-linear model for ENSO variability in terms of global mean surface temperature. Panel a shows currents as a function of latitude and depth (averaged over the Pacific basin, 140-270$^\circ E$), with the characteristic latitude of descent (solid) and rough uncertainty range (dashed) indicated with vertical lines. Panel b shows the year after warming begins that temperature maxima (red), temperature minima (blue), and temperature at depth (green) reach half of their final warming as a function of surface warming timescale (uncertainty given as the ensemble range). The solid black line represents no delay between surface warming and the quantity in question, the dashed black lines represent delays in intervals of 50 years, the solid green line represents a delay of $t_{\mathrm{STC}}=36$ years, and the dashed green lines represent an uncertainty around this value (calculated from the range of downwelling latitudes). Below panels a and b is the lag-linear model equation. The performance of the lag-linear model at predicting ENSO changes in CanESM5 is shown in panel c, and the relative accuracy of simulated ENSO and the lag-linear model as a function of realizations used is demonstrated in panel d ($R^2$ calculated as a comparison against the full ensemble mean ENSO time series). The zero realization ENSO prediction is calculated using a three-box energy balance model for GMST (see methods section for details).}\label{Fig:LagLinearDerivationFigure}
\end{figure}

Building on Eq. \ref{eq:StratLinearModel}, we develop a lag-linear model for ENSO variability in terms of global mean SST (GMST) by connecting stratification changes to subsurface warming and thereby SST changes with a lag (Fig. \ref{Fig:LagLinearDerivationFigure}). In the control climate, easterly trade winds and the Coriolis effect cause near-surface poleward currents throughout the tropics. The mass moving away from the equator is replaced by strong upwelling, and water returns from the subtropics to the equator at depth \citep[Fig. \ref{Fig:LagLinearDerivationFigure}a,][]{mccreary_interaction_1994,liu_gcm_1994}. This circulation, or the ocean's ``subtropical cells," controls the timescale over which subsurface water on the equator responds to greenhouse warming.

By diagnosing a simulation's subtropical cells, we can calculate an expected replacement time of equatorial subsurface water. Dividing the southernmost latitude of climatological downwelling (estimated at $\phi_\downarrow \approx 16^\circ$) by a representative southward velocity ($v_{\mathrm{STC}} \approx 1.6$ mm/s, averaged over the top 150m and calculated as the harmonic mean with respect to latitude because transit time is inversely proportional to velocity) gives an expected subtropical cell timescale, or how long it takes for surface water to replace equatorial subsurface water ($t_{\text{STC}} = \phi_\downarrow/v_{\mathrm{STC}}$= 36 years). We assume the two hemispheres have similar subtropical cells; this is likely as we are studying an ocean region with no hemispherically asymmetric continents and the trade winds are centered near the equator.

The timescale of the subtropical cells matches the lag between rising El Ni\~no and La Ni\~na temperatures (Fig. \ref{Fig:LagLinearDerivationFigure}b). As discussed previously, temperature maxima increase at almost exactly the same rate as regional surface temperatures, while temperature minima, as well as subsurface water, warm several decades later. This lag matches the diagnosed $t_{\text{STC}}$ based on the simulations' currents and, crucially, indicates that the delayed response of warming at depth does not depend on the warming timescale. 

It is now possible to predict ENSO strength from surface temperature and its history using the connection between temperature at depth and subtropical surface temperatures with a lag. We assume that surface warming in each region (i.e., the equatorial East Pacific and the subtropical Pacific) is proportional to GMST increases and that East Pacific mixed layer temperatures are proportional to the average of East Pacific surface and subsurface temperatures (Supplementary Text 3), leading to a lag-linear model of the form:

\begin{equation}
\label{eq:Lag-linear}
    \Delta \mathrm{M}_{\mathrm{ENSO}}(t)=\alpha \Delta\overline{\mathrm{GMST}}(t) + \beta \Delta \overline{\mathrm{GMST}}(t-t_{\mathrm{lag}})
\end{equation}
where $\Delta\overline{\mathrm{GMST}}$ is the global mean SST change from a control climate (smoothed by a 20-year moving mean as with other bar variables), $t_{\text{lag}}$ is a timescale which represents how long it takes for warming to reach the subsurface in the East Pacific, and $\alpha>0$ and $\beta<0$ are constants (quantitatively derived from $A$ and $B$ in Supplementary Text 3).

The lag-linear model can accurately predict ENSO variability in a wide range of simulations. With the same parameters across warming scenarios (but differing across models), it captures nearly all of the ENSO signal in MITgcm, CanESM5, CESM2, and EC-Earth3 ensembles, while performing reasonably well in MIROC6 (Fig. \ref{Fig:LagLinearDerivationFigure}c and Extended Data Fig. \ref{Fig:ED_SSPsLagLinear}). Additionally, it replicates changes in ENSO variability under an abrupt quadrupling of CO$_2$ in most CMIP models and the ensemble mean (Extended Data Fig. \ref{Fig:ED_AbruptLagLinear}). Beyond these transient warming cases, the lag-linear model, with coefficients fit to the $\Delta t=50$ yr simulation, can predict steady-state ENSO variability in a range of climates simulated by the MITgcm, with $R^2>0.98$ (Extended Data Fig. \ref{Fig:ED_Paleo}a). In CESM1 simulations representing climates ranging from the last glacial maximum to the present, the lag-linear model correctly predicts that ENSO variability is proportional to GMST over the past 12,000 years. Before then, however, factors such as changing land-masses and orbital parameters alter ENSO variability in ways not captured by the lag-linear model (Extended Data Fig. \ref{Fig:ED_Paleo}b). A full list of all lag-linear fits, their parameter values, and their $R^2$ values is displayed in Extended Data Table 2.

%The lag-linear model, using the same values of $\alpha$, $\beta$, and $t_{\mathrm{lag}}$, accurately predicts ENSO variability across MITgcm warming scenarios (Fig. \ref{Fig:IntroFigure}f, Extended Data Fig. \ref{Fig:ED_SSPsLagLinear}a, and Extended Data Table 2). Additionally, the vital time lag parameter is possible to diagnose from the ocean circulation; the fit $t_{\text{lag}}$ in the MITgcm is about 34.6 years, very close to $t_{\text{STC}}=35$ years. Similarly, the lag-linear model (with different parameters) works in CanESM5 (Fig. \ref{Fig:IntroFigure}c and Fig. \ref{Fig:LagLinearDerivationFigure}, $R^2=0.95$) and several other CMIP models under SSP emissions (Extended Data Fig. \ref{Fig:ED_SSPsLagLinear} and Extended Data Table 2, with the exception of MIROC6), in addition to abrupt quadrupling CO$_2$ scenarios from individual models (Extended Data Fig. \ref{Fig:ED_AbruptLagLinear}, Extended Data Table 2, and Fig. \ref{Fig:IntroFigure}d). Lastly, the lag-linear model correctly predicts that ENSO variability should be linear in GMST across a range of paleo-climate simulations using CESM1 (until we go back far enough that orbital parameters and sea-level have changed significantly, Extended Data Fig. \ref{Fig:ED_Paleo}), and the lag-linear model with parameter values diagnosed from the $\Delta t=50$ simulations can predict the MITgcm steady-state ENSO variability in a range of climates. A full list of all lag-linear fits, their parameter values, and their $R^2$ values is displayed in Extended Data Table 2.

The lag-linear model provides reliable predictions of ENSO variability with little computational cost. While directly diagnosing $\Delta \mathrm{M}_{\mathrm{ENSO}}$ typically requires at least 10-15 realizations to reasonably match a large ensemble \citep{maher_future_2023}, the lag-linear model requires far fewer (shown for CanESM5 in Fig. \ref{Fig:LagLinearDerivationFigure}d and other models in Extended Data Table 3). Because GMST generally has less internal variability than ENSO, the lag-linear predictor can match an ensemble mean prediction with minimal realizations (Fig. \ref{Fig:LagLinearDerivationFigure}d). In fact, by predicting GMST using a three-box energy balance model such as the Finite Amplitude Impulse Response Simple Climate Model (FaIR) \citep[see methods section,][]{leach2021fairv2}, we can accurately predict ENSO variability without running a GCM at all (Fig. \ref{Fig:LagLinearDerivationFigure}d and Extended Data Table 3).

The values of $\alpha$ and $\beta$ vary considerably across models mostly due to differing simulated ENSO responses to warming. As GCMs predict different responses of ENSO to the same emissions scenario \citep[][and Extended Data Fig. \ref{Fig:ED_AbruptLagLinear}d]{cai_increased_2018,callahan_persistent_2023}, our linear predictors must use different parameter values to approximate each model. Despite quantitative inter-model disagreement in the parameters, the accuracy of the lag-linear predictor is robust across nearly all studied cases.

\subsection*{Solving for Peak ENSO Magnitude}

In addition to being an efficient predictor, the lag-linear model gives insight into what controls the amplitude and timing of peak ENSO variability in a warming climate. The parameters $\alpha$ and $\beta$ represent the sensitivity of ENSO to changes in surface and subsurface temperature, respectively (discussed in Supplementary Text 3), while $t_\mathrm{lag}$ corresponds to the subtropical cell timescale and can be diagnosed from simulated currents (MITgcm fit $t_{\text{lag}}=$ 34.6 yr, close to $t_{\text{STC}}=36$ years). To derive an analytic solution for ENSO variability, we assume GMST follows idealized trajectories of the form:

\begin{equation}
\label{eq:TempEquation}
\Delta \overline{\mathrm{GMST}}=\frac{\Delta T_{\infty}}{2} \left(1+\mathrm{erf} \left(\frac{t-t_0}{\tau_W} \right) \right)
\end{equation}
where the three parameters are $\Delta T_{\infty}$ representing the warming amplitude, $\tau_W$ representing a warming timescale, and $t_0$ representing the time of the fastest warming. This form, with only three degrees of freedom, can reproduce the warming signal in full CESM2 simulations (time series in Fig. S6). 

\begin{figure}[h]
\centering
\includegraphics[width=0.9\textwidth]{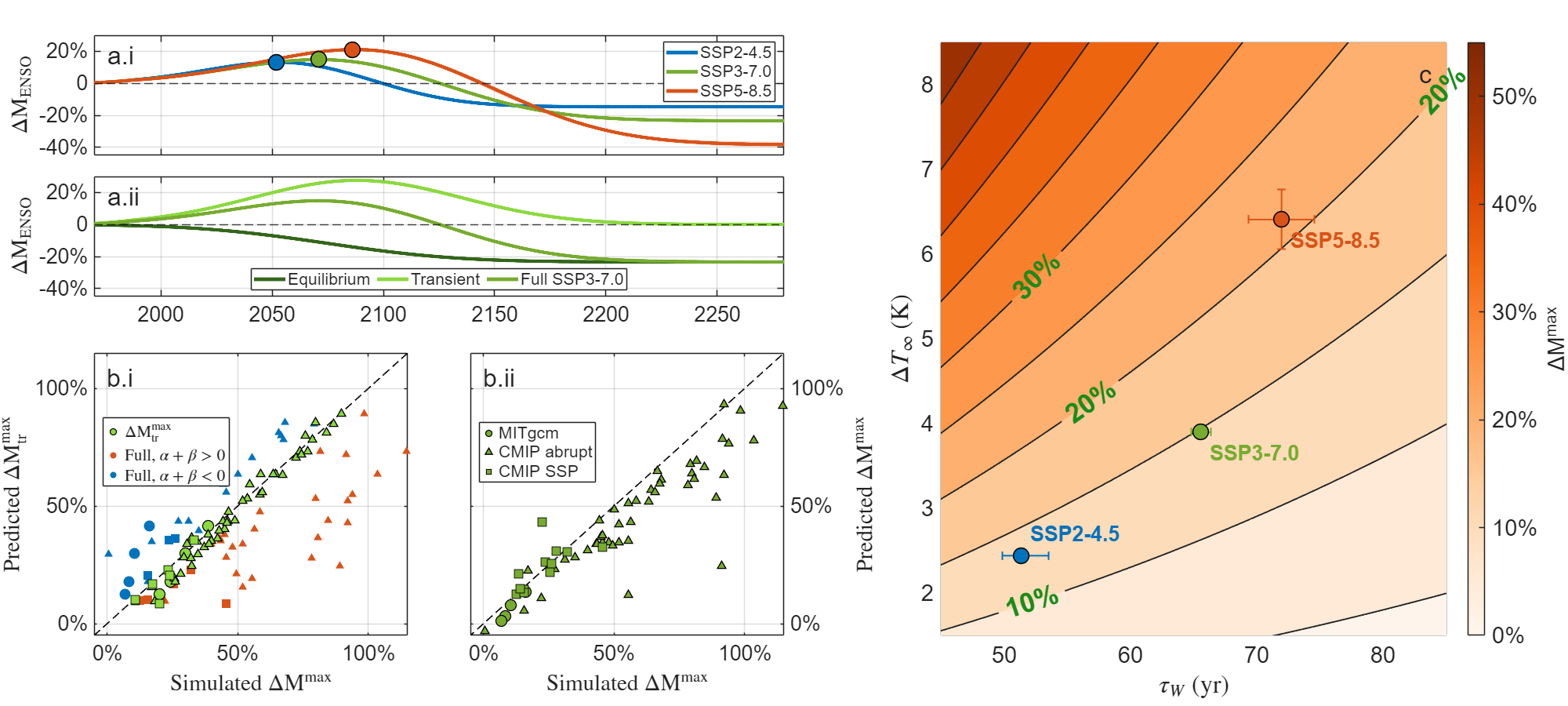}
\caption{Predictions from the lag-linear model and idealized temperature trajectory. Panel a shows ENSO variability over time, with a.i displaying three common SSPs (using CESM2 fits) and a.ii splits the SSP3-7.0 response into an equilibrium and transient response. Panel b.i shows the maximum amplitude of the diagnosed transient component (green) and the maximum amplitude of the full simulation vs. the predicted peak of the transient component. The full peaks are split into those simulations which predict weaker ENSO events in a warmer climate (blue) and those that predict stronger ENSO events in a warmer climate (red). Panel b.ii shows the full simulated ENSO peaks against the full predicted ENSO peak using the expressions derived in Supplemental Text 4. Finally, panel c shows the maximum amplitude of ENSO events as a function of warming timescale and warming amplitude. The error bars on the SSP scenarios are calculated from an n=100 bootstrap of the temperature trajectory fits.}\label{Fig:LagLinearPredictionsFigure}
\end{figure}

Using the idealized GMST projection, the lag-linear model can estimate ENSO variability past 2100 from a wide range of emission pathways (Fig. \ref{Fig:LagLinearPredictionsFigure}a.i). In each scenario, ENSO variability first rises and then falls to well below the preindustrial value. The SSP2-4.5 scenario (``Middle of the Road") shows ENSO variability reaching up to 15\% higher than preindustrial levels and remaining elevated through about 2100.  Interestingly, ENSO variability peaks only slightly higher in the SSP3-7.0 scenario (``Regional Rivalry"), but remains elevated until about 2125. In the SSP5-8.5 scenario (``Fossil-fueled Development"), ENSO variability reaches about 20\% higher than preindustrial levels and remains elevated well into the 22nd century. 

Once a simplified temperature trajectory is assumed, six parameters control the behavior of ENSO variability, each of which has units of temperature or time: $\alpha$, $\beta$, $t_{\mathrm{lag}}$, $\Delta T_{\infty}$, $\tau_W$, and $t_0$. The parameter representing when warming occurs, $t_0$, can alter peak timing but not amplitude, leaving three non-dimensional parameters that control the magnitude of peak variability: $\alpha \Delta T_{\infty}$, $-\beta/\alpha$, and $t_{\mathrm{lag}}/\tau_W$ (discussed below and in Supplementary Text 4).  

\begin{comment}
We can better predict and understand the timing and amplitude of peak ENSO variability by separating $\Delta \mathrm{M}_{\mathrm{ENSO}}$ into an equilibrium component:

\begin{equation}
    \Delta \mathrm{M}_{\mathrm{eq}}\equiv(\alpha+\beta)\Delta \overline{\mathrm{GMST}}
\end{equation}
or ENSO strength if the system came to equilibrium at the current temperature, and a transient component:
\begin{equation}
    \Delta \mathrm{M}_{\mathrm{tr}}=-\beta \left(\Delta \overline{\mathrm{GMST}}(t)-\Delta \overline{\mathrm{GMST}}(t-t_{\mathrm{lag}}) \right),
\end{equation}
decomposition shown in Fig. \ref{Fig:LagLinearPredictionsFigure}a.ii. 

\end{comment}

We can make progress towards understanding the amplitude and timing of peak ENSO variability by separating $\Delta \mathrm{M}_{\mathrm{ENSO}}$ into an equilibrium component representing steady-state ENSO variability at a given GMST and a transient component representing the increase in $\Delta \mathrm{M}_{\mathrm{ENSO}}$ from differing surface and subsurface changes (Fig. \ref{Fig:LagLinearPredictionsFigure}a.ii):
\begin{equation}
\Delta \mathrm{M}_{\mathrm{ENSO}}(t) \;=\; \underbrace{(\alpha+\beta)\,\Delta \overline{\mathrm{GMST}}(t)}_{\Delta \mathrm{M}_\mathrm{eq}(t)} \;+\; \underbrace{-\beta\bigl[\Delta \overline{\mathrm{GMST}}(t)-\Delta \overline{\mathrm{GMST}} (t-t_{\mathrm{lag}})\bigr]}_{\Delta \mathrm{M}_\mathrm{tr}(t)},
\label{eq:decomposition}
\end{equation}

The transient component ($\Delta \mathrm{M}_\mathrm{tr}$) is controlled by the change in temperature over the last $t_{\mathrm{lag}}$, so its peak is set by the maximum warming that occurs in a fixed period. The amplitude of the strongest ENSO events therefore depends on both a change in temperature and a warming timescale (i.e., $\tau_W$ compared to $t_\mathrm{lag}$). Using the error function form of $\Delta \overline{\mathrm{GMST}}(t)$ (calculations shown in Supplementary Text 4), the maximum of the transient component is:

\begin{equation}
    \Delta \mathrm{M}_{\mathrm{tr}}^{\mathrm{max}} =-\beta \Delta T_{\infty}\mathrm{erf} \left(\frac{t_{\mathrm{lag}}}{2\tau_W} \right).
    \label{eq:TransientPrediction}
\end{equation}
This expression agrees well with a wide range of simulation results (Fig. \ref{Fig:LagLinearPredictionsFigure}b.i), and tells us that peak ENSO variability increases with warming amplitude and warming speed relative to the constant timescale $t_{\mathrm{lag}}$.  The peak is predicted to occur at $t_0+t_{\mathrm{lag}}/2$, or half a subtropical cell timescale after the time of maximum warming. We use a different form of GMST trajectory in the case of abrupt warming, but the system can still be solved analytically and the physical insights remain robust (see Supplementary Text 4).

The equilibrium component ($\Delta \mathrm{M}_\mathrm{eq}$), meanwhile, alters peak ENSO variability according to the long-term response of ENSO strength to warming. Because $\Delta \overline{\mathrm{GMST}}$ is uniformly positive and increasing, the sign of the equilibrium component is constant and the same as that of $\alpha+\beta$. The sign of $\alpha+\beta$ also represents the predicted steady-state response of ENSO in a warmer climate; as $t\to\infty$, $\Delta \mathrm{M}_{\mathrm{ENSO}} \to (\alpha+\beta)\Delta T_{\infty}$ (as long as $\Delta \overline{\mathrm{GMST}}$ approaches a constant value). This means that if the model predicts less ENSO variability in warmer climates, i.e., $\alpha+\beta<0$, then $\Delta \mathrm{M}_{\mathrm{eq}}<0$ and peak ENSO variability will be shifted earlier and smaller. If, on the other hand, the model predicts more ENSO variability in warmer climates, then $\alpha+\beta>0$ and peak variability will be shifted later and larger. This heuristic, while simple, is very accurate across all simulations studied (Fig. \ref{Fig:LagLinearPredictionsFigure}b.i). 

The error function form of temperature trajectories allows for a full solution of peak time and magnitude (Supplementary Text 4), and the resulting predictions match simulated peaks well (Fig. \ref{Fig:LagLinearPredictionsFigure}b.ii, different temperature trajectory form used for abrupt warming simulations). The key non-dimensional parameters are a normalized warming amplitude ($\alpha \Delta T_{\infty}$), an inverse relative warming timescale ($t_\mathrm{lag}/\tau_W$), and a measure of the relative sensitivity of ENSO to surface and subsurface warming ($-\beta/\alpha$); this is discussed in Supplementary Text 4 and shown in Extended Data Fig. \ref{Fig:ED_Analytic}. 

The full solution can be used to predict quantitatively how peak ENSO variability depends on warming timescale and amplitude (Fig. \ref{Fig:LagLinearPredictionsFigure}c). Unsurprisingly, stronger warming scenarios (i.e., larger $\Delta T_{\infty}$) at a constant warming timescale lead to larger peak amplitudes. Much more interestingly, peak ENSO amplitude also depends strongly on the warming timescale $\tau_W$; shorter timescales (faster emissions) lead to larger ENSO peaks even if the total emissions are the same. This explains the predictions from SSP2-4.5 and SSP3-7.0 in panel a.i; while SSP3-7.0 has significantly more warming ($\Delta T_{\infty}\sim$4 K vs. 2.5 K), the warming occurs over a longer timescale ($\tau_W\sim$65 years vs. 50), so the peak ENSO amplitude is similar. Intuitively, if subsurface warming can keep pace with surface warming, then peak ENSO amplitude is limited, while if significant warming occurs suddenly, ENSO variability increases drastically.

\section*{Discussion}\label{sec:Discussion}

El Ni\~no and La Ni\~na events control a large part of climate's impact on society by altering large-scale circulations and global weather patterns. The strength of these events, or ENSO variability, is predicted to undergo a transient rise and permanent fall under greenhouse warming. Here, we have shown that the temporary enhancement of ENSO strength comes from an immediate increase in surface temperatures, while the subsequent reduction in ENSO strength is associated with 1. decreasing stratification due to the warming subsurface and 2. a slower Walker circulation and stronger surface flux damping. We use this understanding to develop two simple linear models which can predict ENSO variability quickly and accurately. The second of these models can be solved analytically for idealized temperature trajectories, revealing that peak ENSO variability depends on warming amplitude, a relative sensitivity of ENSO to surface and subsurface temperature changes, and the timescale of warming compared to a subsurface ocean adjustment timescale. We find that fast warming will lead to very strong ENSO events before their eventual weakening, while slower warming moderates this peak and prevents the most extreme impacts. This implies that a given amount of greenhouse gas emissions can have different effects at different times, i.e., the social cost of carbon is time-dependent.

There are limits to the accuracy of the lag-linear model given our current state of knowledge about the tropical Pacific. Specifically, the tendency of models to underestimate short term warming in the equatorial East Pacific will likely lead to biases in the diagnosed value of $\alpha$, $\beta$, and $t_{\mathrm{lag}}$. Similarly, simplifications in our idealized MITgcm simulations, such as a lack of anomalous shortwave radiation, are likely to cause quantitative errors in the linear model parameters (i.e., $A$ and $\alpha$).

The biases leading to these errors, in particular those in simulating the East Pacific under warming, may be ameliorated by applying our framework for studying the differing response timescales of the surface and subsurface ocean. The climatological East Pacific temperature is influenced by upwelling just as much as ENSO, so delayed subsurface warming is critical for understanding the region's response to climate change \citep{clement1996ocean}. Future research focused on errors in the subtropical cell timescale may reveal why the East Pacific experiences too much short-term warming in simulations \citep{vecchi_examining_2008,seager_strengthening_2019} and help us understand the well-known ``pattern effect" \citep{zhou2021greater,dong2020intermodel}.

\begin{comment}
Until the pattern effect is understood, however, it may cause uncertainties in the diagnosed values of $\alpha$, $\beta$, and $t_{\mathrm{lag}}$. If the models have significant biases in how the East Pacific responds to warming, it is likely that their diagnosed or fitted $t_{\mathrm{lag}}$ will not match that of reality. It is also worth noting that simplifications in our idealized MITgcm simulations, such as a lack of anomalous shortwave radiation, are likely to cause quantitative errors in the linear model parameters (i.e., $A$ and $\alpha$).
\end{comment}

With respect to broader climate research, the utility of our linear models indicates that simple predictors could be used to study a wide range of phenomena across different regimes. For example, ENSO in paleoclimate contexts may depend on relatively few variables such as basin width and mean temperature, and might therefore be explainable with simple models. Similarly, other climate modes (such as the North Atlantic Oscillation or Pacific Decadal Oscillation) may change under greenhouse warming mostly in proportion to mean temperature and one or two other variables, giving the techniques used here broad applicability.

Finally, our quick and accurate lag-linear model facilitates the prediction of ENSO impacts in a wide range of scenarios across which it would be computationally impossible to run a CMIP ensemble. To reliably diagnose ENSO strength over time from Earth system models requires at least a dozen realizations from each of several models \citep{maher_future_2023}, and so is often prohibitively expensive. Forecasts of phenomena affected by ENSO need to consider how El Ni\~no events will evolve over time; using our lag-linear model as a proxy for future ENSO variability can accelerate research on topics ranging from monsoon variability and tropical cyclones to the economic consequences of climate change \citep{callahan_persistent_2023}. Overall, the societal impact of a warming climate will depend strongly on the evolution of ENSO variability; a fast prediction method for this crucial quantity will allow for a more accurate calculation of the social cost of carbon across emission scenarios and help society better prepare for future climate extremes.

\clearpage

\section*{Methods}\label{sec:Methods}
\renewcommand{\thefigure}{M.\arabic{figure}}
\setcounter{figure}{0}

We now give details on the methods used in this work, beginning with a description of the comprehensive and idealized climate models, followed by a discussion of predicting ENSO variability from ocean energy balance, and concluding with an explanation of how we use an energy balance model to forecast GMST from a given emissions scenario. 

\subsection*{Comprehensive Models} 
\label{subsec:MethodsCMIP}

Our most accurate predictions of future ENSO variability come from models participating in the Sixth Coupled Model Intercomparison Project \citep[CMIP6,][]{eyring_overview_2016}. We use four models to simulate realistic warming scenarios (CanESM5, CESM2, EC-Earth3, and MIROC6), with the number of realizations available from each shown in Extended Data Table 2. For three of these models SSPs 1, 2, 3, and 5 are used, while for CESM2 only SSPs 2, 3, and 5 are available. Time series from the historical run and each emissions scenario are concatenated before averaging across realizations. We also use one realization from each of 46 models undergoing an abrupt quadrupling of CO$_2$, listed in Extended Data Table 2. For models that publish the necessary data, the piControl and abrupt4xCO$_2$ simulations are concatenated to form a continuous time series; if that data is not available, then the abrupt4xCO$_2$ is simply concatenated to the end of the piControl run. This does not lead to a significant discontinuity because the internal variability is typically smaller than the forced response -- and if that is not the case then the lag-linear model is unlikely to fit well regardless (Extended Data Fig. \ref{Fig:ED_AbruptLagLinear}f). For the abrupt simulations, the lag-linear fit is conducted over the 80 years after forcing is applied, except in Extended Data Fig. \ref{Fig:ED_UnsmoothedENSO} which uses 60 years to avoid the noise after that.

\begin{comment}
We use CESM2 to understand how ENSO responds to commonly used SSP emission scenarios; all quantities displayed for the SSP3-7.0 scenario (Fig. \ref{Fig:IntroFigure}) are the mean across 100 realizations from the CESM2 Large Ensemble Community Project \citep{rodgers_ubiquity_2021}, while all quantities for the SSP2-4.5 and SSP5-8.5 scenarios (Fig. \ref{Fig:IntroFigure}f and Fig. \ref{Fig:TimescaleFigure}) are the average across 15 realizations. For each SSP, time series from the historical run and that emissions scenario are concatenated before averaging across realizations. Additionally, when fitting temperature curves to each warming scenario (Extended Data Fig. \ref{ed:TempTrajectories}), we employ temperature predictions at 2300 from \citet[][using the MAGICC7 emulator]{lee2021future}. These long-term predictions are included in the fitting process and their uncertainties are taken to be the uncertainty in $\Delta T_{\infty}$. 
\end{comment}

For all CMIP data, sea surface temperatures (variable name \texttt{tos}) averaged from 210-270$^\circ$ longitude within 5 degrees of the equator represent the East Pacific (Ni\~no3) while data averaged from 130-290$^\circ$ longitude within 5 degrees of the equator represents the tropical Pacific (for Extended Data Fig. \ref{Fig:ED_ElNinoVsLaNina}b).

Two of the CMIP models require special treatment due to data availability. NorCPM1 runs only 80 years after the abrupt quadrupling of CO$_2$, so it is padded with the mean of the final 30 years (it is close to equilibrium by then and the fit does not include that data). EC-Earth3 data is normalized by only 10 years (1970-1980) because historical simulation output is not available before that. 

\subsection*{MITgcm Simulations}
\label{subsec:MethodsMITgcm}

To study ENSO in an idealized setting, we run simplified MITgcm simulations \citep{marshall_finite-volume_1997} in configurations based on those in \citet{tuckman_enso_2025}; the code for the control run is available at \url{https://github.com/MITgcm/verification_other/tree/master/cpl_gray%2Bswamp%2Bocn}. The model has a West Pacific warm pool and East Pacific cold tongue comparable to those in observations, as well as tropical easterlies, extratropical westerlies, and an ENSO mode with approximately correct spatial and temporal patterns \citep{tuckman_zonal_2024,tuckman_enso_2025}.

The MITgcm simulations use cubed-sphere grids with $\sim$2.8$^\circ$ resolution in the tropics for the atmosphere and ocean \citep{adcroft_implementation_2004}, have 26 vertical levels in the atmosphere, and have 43 vertical levels in the ocean (with more levels near the surface). The atmosphere uses idealized moist physics and a gray radiation scheme \citep{frierson_gray-radiation_2007} which includes water vapor feedback on long-wave optical thickness \citep{byrne_landocean_2013} but does not include clouds or shortwave absorption by the atmosphere. There is a seasonal cycle of incident shortwave radiation appropriate for a circular orbit with an obliquity of 23.45$^\circ$. The ocean has a uniform depth of 3.4 km. 

The model's continental configuration consists of a large landmass meant to represent Eurasia which extends from 0-135$^\circ$E and from $\sim8^\circ$N to the North Pole. This landmass is treated as a 2m slab ocean so that it has a low heat capacity and no ocean dynamics. Other continents are represented as thin barriers that block ocean flow: two barriers at 0 and 270$^\circ$E extend from 35$^\circ$S to the North Pole to isolate the Atlantic basin, and one extends from 30$^\circ$S to the southeast corner of the continent to separate the Indian and Pacific basins. 

The MITgcm is run for 750 years in a control, preindustrial-like climate, then a time-stepping parameter ($\texttt{abEps}$ in the code) is branched to slightly different values in order to create different realizations without changing the physics of the model. Forty years after the time-stepping parameter is modified, at a time defined to be $t=0$ years, greenhouse warming begins through an increase in the longwave absorption efficiency of CO$_2$ ($\tau_{CO_2}$ or $\texttt{ir\_tau\_co\_2}$ in the code). Over the following $\Delta t$ (10, 50, 100, or 150 years), $\tau_{CO_2}$ increases linearly from 0.8678 to 1.3017. In addition to one control run in which $\tau_{CO_2}$ is not changed, there are 25 realizations used for each of the four values of $\Delta t$.

\subsection*{ENSO Magnitude Prediction from Energy Balance}
\label{subsec:MethodsENSO}

Here, we explain in detail the energy balance framework used to study how and why the magnitude of ENSO events changes as the climate warms. This method is similar to the Bjerknes index \citep{jin_coupled-stability_2006} in that it predicts how ENSO variability changes from mean variable evolution. Unlike the Bjerknes index, our method 1. evaluates products before taking volume averages, 2. includes anomalous meridional advection, 3. makes no assumptions about the meridional structure of temperature anomalies, and 4. allows relationships between anomalous variables to depend on $\tau$. Both methods assume a constant structure of ENSO as the climate warms and treat surface fluxes as a linear damping term. We begin with a general ocean energy budget then apply it to predicting ENSO variability.

Temperature tendency in the ocean is given by:

\begin{equation*}
    \frac{\partial}{\partial t} T(\mathbf{x},t) = - \mathbf{\nabla} \cdot \left(\mathbf{u} T\right) + \frac{\partial}{\partial z} F_{\text{SGS}}
\end{equation*}
where $T$ is temperature as a function of space ($\mathbf{x}$) and time (t), $\mathbf{u}$ is the 3D-current, $\nabla \cdot$ is the 3D divergence, and $F_{\text{SGS}}$ represents any sub-grid-scale or diabatic process including mixing, radiation, and turbulent fluxes (i.e., sensible heat and evaporation). It is assumed that the sub-grid-scale forcing acts only vertically. 

As we wish to study anomalous temperatures, each variable (e.g., $T$) is separated into a 20-year causal moving mean ($\overline{T}$), a mean seasonal cycle ($\tilde{T}$), and an anomaly ($T'=T - \overline{T}-\tilde{T}$). The anomalous temperature budget is:

%The 20-year moving mean is calculated so that it includes the seasonal cycle, i.e., the value of $\bar{T}$ in January of year $n$ is the average of Januaries from year $n-10$ to $n+10$. 

\begin{equation*}
    \frac{\partial}{\partial t} T'(\mathbf{x},t) = - \left(\mathbf{\nabla} \cdot \mathbf{u}T \right)' + \frac{\partial}{\partial z} F_{\text{SGS}}'.
\end{equation*}

We now decompose the flux-form advection term into an advective-form term and a flux-form correction $\text{FF}_{\text{corr}}$ defined such that $\left(\mathbf{\nabla} \cdot \mathbf{u}T\right)' = \left(\mathbf{u} \cdot \nabla T \right)'+\text{FF}_{\text{corr}}'$. As seawater is effectively incompressible, $\text{FF}_{\text{corr}}'$ should be zero, but because data for $\mathbf{u}$ and $T$ are evaluated only once a month, $\text{FF}_{\text{corr}}'$ has small, finite values corresponding to short timescale correlations. The anomalous advection term can now be written out as:
\begin{equation}
    \left(\mathbf{u} \cdot \nabla T \right)' = \overline{\mathbf{u}} \cdot \nabla T' + \tilde{\mathbf{u}} \cdot \nabla T' + \mathbf{u}' \cdot \nabla \overline{T} +  \mathbf{u}' \cdot \nabla \tilde{T} + \left(\mathbf{u}' \cdot \nabla T' \right)'.
\end{equation}

To simplify the expression of small terms that are not of interest, we define a seasonal cycle advection term such that $\text{Adv}_{\text{SC}}'$ ($\equiv-\tilde{\mathbf{u}} \cdot \nabla T' - \mathbf{u}' \cdot \nabla \tilde{T}$). This makes the anomalous energy budget:

\begin{equation*}
    \frac{\partial}{\partial t} T'(\mathbf{x},t) = -\mathbf{u}' \cdot \nabla \overline{T} -\overline{\mathbf{u}} \cdot \nabla T' - \left(\mathbf{u}' \cdot \nabla T' \right)' +\text{Adv}_{\text{SC}}'+ \text{FF}_{\text{corr}}' + \frac{\partial}{\partial z} F_{\text{SGS}}'.
\end{equation*}

Next, we take the volume integral over the equatorial East Pacific upper ocean, defined as the region between $\sim210$ and $270^\circ$E, within $\sim5$ degrees of the equator (corresponding to Ni\~no3), and the top $\sim$80 m. These bounds are approximate as the spatial integration is conducted on the native cubed sphere grid. Denoting the volume average of a variable as $\langle \cdot \rangle=\int \int \int \cdot dx dy dz/(\Delta_x \Delta_y \Delta_z)$, where $\Delta_x, \Delta_y$, and $\Delta_z$ correspond to the zonal, meridional, and vertical distance across the box studied:

\begin{equation*}
    \frac{d}{d t} \langle T' \rangle = -\langle \overline{\mathbf{u}} \cdot \nabla T' \rangle - \langle \mathbf{u}' \cdot \nabla \overline{T} \rangle - \langle (\mathbf{u}' \cdot \nabla T')'  \rangle + \langle \text{Adv}_{\text{SC}}' + \text{FF}_{\text{corr}}' \rangle + \left \langle \frac{\partial}{\partial z} F'_{\text{SGS}} \right\rangle.
\end{equation*}
For the SGS forcings, we evaluate the vertical integral, so $\langle \partial F'_{\text{SGS}}/\partial z \rangle=\langle F' \rangle_{\mathrm{2D,surf}} + \mathrm{\it{Diffusion}}$, where $\langle \cdot \rangle_{\mathrm{2D,surf}}$ indicates the area average at the ocean's surface and $F'=\text{SH}'+\text{LH}'+\text{LWR}'$ represents the sum of sensible heat, latent heat, and longwave radiation (shortwave radiation has no anomalous contribution in the MITgcm, all terms in K/s). From here, diffusion will be treated as part of the flux form correction (it does not affect ENSO significantly, shown below). Note that the mixed layer depth could change, but we find that the variability of the studied box is a very good predictor of surface temperature variability over time.

% and neglect mixing at the bottom (due to the relatively slow speed of interior ocean mixing \citep{munk1966abyssal})%

As we wish to understand how the mean climate and anomalous temperatures interact, we assume that net anomalous surface fluxes (including radiation) are proportional to anomalous surface temperature via a slowly changing constant $\bar{\alpha}_{\mathrm{SF}}$. This assumption is equivalent to treating surface fluxes as a damping term for ENSO anomalies, as is often done when analyzing ENSO energy budgets \citep[][]{jin_coupled-stability_2006,jin_simple_2020,tuckman_understanding_2025}. The budget is now:

\begin{equation}
\label{eq:MethodsBudget}
    \frac{d}{d t} \langle T' \rangle = -\langle \overline{\mathbf{u}} \cdot \nabla T' \rangle - \langle \mathbf{u}' \cdot \nabla \overline{T} \rangle - \langle (\mathbf{u}' \cdot \nabla T')'  \rangle + \langle \text{Adv}_{\text{SC}}' +  \text{FF}_{\text{corr}}' \rangle - \bar{\alpha}_{\mathrm{SF}} \langle T' \rangle_{\text{2D,surf}}
\end{equation}
where $\bar{\alpha}_{\mathrm{SF}}$ is calculated as the proportionality constant between $\langle T' \rangle_{\text{2D,surf}}$ and $\langle F' \rangle_{\text{2D,surf}}$ across positive values of $\langle T' \rangle_{\text{2D,surf}}$ within 10 years of time $t$ (see Supplementary Text 1 for discussion of this assumption and how $\bar{\alpha}_{\mathrm{SF}}$ changes with warming). 

We test our budget by applying Eq. \ref{eq:MethodsBudget} to a random time interval in Fig. S7a. The budget for this sample time series is mostly closed, and across all time points the budget prediction and simulated tendency are well correlated (panel b). There are residuals at some times, likely associated with model output being monthly.

We can also apply East Pacific energy balance to a composite El Ni\~no event in order to understand how ENSO events grow. We create a composite by first identifying the 25 largest values of $\langle T' \rangle$ divided by the moving East Pacific standard deviation. This normalization is necessary so that the peaks identified are spread out across climates, rather than concentrated in climates with larger ENSO variability. Large values within two years of each other are ignored. The identified events are then composited according to a time relative to the El Ni\~no peak $\tau$ (defined such that the maximum $\langle T' \rangle$ is at $\tau=0$). In other words, a variable at a given $\tau$ represents the average of that quantity across a set of El Ni\~no events a certain number of months before those events peak. The resulting composited time series from the reference simulation and the relevant energy budget terms are shown in Fig. S7c; the budget predicted tendency  (solid black line) and the simulated tendency (dashed black line) match well. To understand El Ni\~no growth we integrate each term over the buildup to the maximum $\langle T' \rangle$ value; temperature increases are dominated by anomalous upwelling acting on mean stratification while sub-grid-scale processes (surface fluxes and radiation) damp temperature anomalies significantly (d). The other terms, especially seasonal advection, diffusion, and the flux form correction, are smaller.

We now have a clear expression for changes to East Pacific anomalous temperature, and we wish to transform it into an expression for ENSO variability. The next step is to divide both sides of the budget by $\langle T' \rangle$:

\begin{equation*}
    \frac{1}{\langle T'\rangle}\frac{d}{d \tau} \langle T' \rangle = \left[-\frac{\langle \overline{\mathbf{u}} \cdot \nabla T' \rangle}{\langle T'\rangle} - \frac{\langle \mathbf{u}' \cdot \nabla \overline{T} \rangle}{\langle T'\rangle} - \frac{\langle (\mathbf{u}' \cdot \nabla T')'  \rangle}{\langle T'\rangle} + \frac{\langle  \text{Adv}_{\text{SC}}' + \text{FF}_{\text{corr}}'\rangle}{\langle T'\rangle} - \bar{\alpha}_{\mathrm{SF}} \frac{\langle T' \rangle_{\text{2D,surf}}}{\langle T'\rangle} \right]
\end{equation*}
so that the tendency contributions include ratios between two anomalous quantities and therefore express the structure of an ENSO event and are not sensitive to the event amplitude. By integrating over time, this budget can be used to diagnose how much $\langle T' \rangle$ will increase in the lead up to an individual El Ni\~no event. We use Simpson's method to integrate from the initiation of an ENSO event at $\tau = \tau_0$ to its peak time $\tau = 0$. The resulting predicted amplitude of an ENSO peak is:

\small
\begin{equation}
\label{eq:BudgetEquation}
    \langle T' \rangle_{\mathrm{pred}} = \langle T' \rangle_{\tau_0} \exp\left[\int_{\tau_0}^0 d\tau\,  \left( - \frac{\langle \mathbf{\bar{u}} \cdot \nabla T' \rangle}{\langle T' \rangle} - \frac{\langle \mathbf{u'} \cdot \nabla \bar{T} \rangle}{\langle T' \rangle} - \frac{\langle \mathbf{u}' \cdot \nabla T'  \rangle}{\langle T'\rangle} + \frac{\langle \text{Adv}_{\text{SC}}' + \text{FF}_{\text{corr}}' \rangle}{\langle T'\rangle} - \bar{\alpha}_{\mathrm{SF}} \frac{\langle T' \rangle_{\mathrm{2D,surf}}}{\langle T' \rangle} \right) \right].
\end{equation}
\normalsize
In order to avoid $\langle T' \rangle = 0$, we define the initiation as the first time when $\langle T' \rangle$ is larger than 0.125 K for composites or 1 K for individual events, which are noisier. It is important to note that predicted ENSO variability is sensitive to this threshold; if it is too high then the initial condition for the integral is biased in warmer climates, while if it is too low the small residual at the beginning of the time series can cause errors. Fig. S8 plots the simulated maximum values of $
\langle T'\rangle$ against those diagnosed by Eq. \ref{eq:BudgetEquation}; the predictions match the simulations reasonably well. 

\subsubsection*{Predicting ENSO Magnitude in a Changing Climate}
\label{subsubsec:ENSOTransfer}

We will now use Eq. \ref{eq:BudgetEquation} to understand how and why El Ni\~no events change with the mean climate (i.e., as $\overline{\mathbf{u}}$, $\nabla \overline{T}$, and $\bar{\alpha}_{\mathrm{SF}}$ evolve). As before, 25 identified El Ni\~no events from the control simulation are used to create composite quantities as a function of time $\tau$. For each event, a time relative to the peak is defined such that $\langle T'\rangle$ is at its maximum value when $\tau=0$, then averages are taken at each value of $\tau$ across the 25 identified events to calculate $\mathbf{u}'(\mathbf{x},\tau)/\langle T'\rangle(\tau)$, $\nabla T'(\mathbf{x},\tau)/\langle T'\rangle(\tau)$, and $\langle T' \rangle_{\text{2D,surf}}(\tau)/\langle T'\rangle(\tau)$. These three variables represent the characteristic structure of an El Ni\~no event and are assumed to be constant with warming (discussed below). We can then use energy balance at any time $t$ to study how $\langle T' \rangle$ grows during an El Ni\~no event in that climate. In other words, we represent a mean climate with $\overline{\mathbf{u}}$, $\nabla \overline{T}$, and $\bar{\alpha}_{\mathrm{SF}}$ as a function of $t$ (averaged over all available realizations) and place a composite El Ni\~no structure represented by $\mathbf{u}'/\langle T'\rangle$, $\nabla T'/\langle T'\rangle$, and $\langle T' \rangle_{\text{2D,surf}}/\langle T'\rangle$ as a function of $\tau$ in that climate. The peak amplitude resulting from an El Ni\~no event in that chosen climate is:

\begin{equation}
    \label{eq:TPrimePrediction}
    \langle T' \rangle_{\mathrm{pred}}(t) = \langle T' \rangle_{\tau_0} \exp\left[\int_{\tau_0}^0 d\tau\,  \left( - \frac{\langle \mathbf{\bar{u}} \cdot \nabla T' \rangle}{\langle T' \rangle} - \frac{\langle \mathbf{u'}_h \cdot \nabla_h \bar{T} \rangle}{\langle T' \rangle} - \frac{\langle w' \bar{\Gamma} \rangle}{\langle T' \rangle} - \bar{\alpha}_{\mathrm{SF}} \frac{\langle T' \rangle_{\mathrm{2D,surf}}}{\langle T' \rangle} \right) \right]
\end{equation}
where $\Gamma \equiv \partial T/\partial z$ is the stratification and the non-linear terms and flux form correction have been ignored here for convenience but are included in the calculations for Fig. \ref{Fig:EnergyBudget}. This is the same equation as above, but now all anomalous variables come from an El Ni\~no composite in the control simulation while the mean variables come from a changing climate. For example, in the stratification term ($\langle w' \bar{\Gamma} \rangle/\langle T' \rangle$) we use the anomalous upwelling per degree anomalous East Pacific temperature ($w'/\langle T' \rangle$) from the reference climate and keep this constant as $t$ changes, but allow the mean stratification ($\bar{\Gamma}$) to evolve. Similarly, for the surface fluxes, $\langle T' \rangle_{\mathrm{2D,surf}}/\langle T' \rangle$ is assumed to be constant as the climate warms, but $\bar{\alpha}_{\mathrm{SF}}$ changes. This equation uses the El Ni\~no event structure from a reference climate and mean climate variables at a given time $t$ to predict the buildup to an El Ni\~no peak if one had happened in that climate. The result is a predicted maximum temperature anomaly as a function of $t$, which changes as the climate warms, i.e., with $\overline{\mathbf{u}}(t)$, $\nabla \overline{T}(t)$, and $\bar{\alpha}_{\mathrm{SF}}(t)$. Additionally, this equation can be evaluated with only one mean variable changing over time, allowing for the calculation of contributions from each term to changes in ENSO peaks (Fig. \ref{Fig:EnergyBudget}d).  

In order to use Eq. \ref{eq:TPrimePrediction} to predict $\Delta \mathrm{M}_{\mathrm{ENSO}}$, we divide both sides by their equivalent reference simulation expression and subtract one, giving:

\begin{equation}
    \frac{\langle T'_{\mathrm{pred}} \rangle}{\langle T'_{\mathrm{pred,cntl}} \rangle}-1 = \frac{\exp\left[\int_{\tau_0}^0 d\tau\,  \left( - \frac{\langle \mathbf{\bar{u}} \cdot \nabla T' \rangle}{\langle T' \rangle} - \frac{\langle \mathbf{u'}_h \cdot \nabla_h \bar{T} \rangle}{\langle T' \rangle} - \frac{\langle w' \bar{\Gamma} \rangle}{\langle T' \rangle} - \bar{\alpha}_{\mathrm{SF}} \frac{\langle T' \rangle_{\mathrm{2D,surf}}}{\langle T' \rangle} \right) \right]}{\exp\left[\int_{\tau_0}^0 d\tau\,  \left( - \frac{\langle \mathbf{\bar{u}}_{\mathrm{cntl}} \cdot \nabla T' \rangle}{\langle T' \rangle} - \frac{\langle \mathbf{u'}_h \cdot \nabla_h \bar{T}_{\mathrm{cntl}} \rangle}{\langle T' \rangle} - \frac{\langle w' \bar{\Gamma}_{\mathrm{cntl}} \rangle}{\langle T' \rangle} - \bar{\alpha}_{\mathrm{SF,cntl}} \frac{\langle T' \rangle_{\mathrm{2D,surf}}}{\langle T' \rangle} \right) \right]}-1.
\end{equation}

Next, we define $\Delta$ terms as differences from control simulation values (e.g., $\Delta \bar{\mathbf{u}}=\bar{\mathbf{u}}-\bar{\mathbf{u}}_{\mathrm{cntl}}$), and the reference terms are moved into the same exponential as the original expression. Lastly, we assume that the normalized ENSO magnitude represented by $\frac{\langle T'_{\mathrm{pred}} \rangle}{\langle T'_{\mathrm{pred,cntl}} \rangle}-1$ is equal to the normalized ENSO magnitude used in the main text (Fig. S1). Our final energy balance prediction is therefore:

\begin{equation}
\label{Eq:MethodsEnergyBalancePrediction}
    \Delta \mathrm{M}_{\mathrm{ENSO}} = \exp \left[ \int_{\tau_0}^0 d \tau \left(- \frac{\langle \Delta \mathbf{\bar{u}} \cdot \nabla T' \rangle}{\langle T' \rangle} - \frac{\langle \mathbf{u}'_h \cdot \Delta \nabla_h \bar{T} \rangle}{\langle T' \rangle} - \frac{\langle w' \Delta \bar{\Gamma} \rangle}{\langle T' \rangle} - \Delta \bar{\alpha}_{\mathrm{SF}} \frac{\langle T' \rangle_{\mathrm{2D,surf}}}{\langle T' \rangle} \right) \right]-1.
\end{equation}

Eq. \ref{Eq:MethodsEnergyBalancePrediction} is the final expression used to predict ENSO magnitude in Fig. \ref{Fig:EnergyBudget} and is a robust and flexible way to understand how changes in climate affect ENSO peaks. However, this method assumes changes to the structure of El Ni\~no events, i.e., the spatial and $\tau$ dependence of quantities such as $w'/\langle T' \rangle$, do not significantly alter the magnitude of ENSO as the climate warms. This is a common assumption in the study of ENSO \citep{jin_coupled-stability_2006,jin_simple_2020,tuckman_enso_2025}, and is necessary due to the difficulty of predicting the precise structure of the anomaly variables over time. We now assess the validity of this assumption by first displaying ENSO structures over time then directly evaluating how changes in structure alter the predicted amplitude.

\subsubsection*{ENSO Structure over Time}
\label{subsubsec:ENSOStructure}

The time-dependent structure of a composite ENSO anomaly is shown in Fig. S9. Although there is some noise, the volume average anomalous quantities as a function of $\tau$ are not significantly different in a warmer climate from those of the reference climate (panel a). Additionally, the spatial structure of normalized anomalous quantities is similar in the control and warmed climate (panel b), although the temperature anomaly becomes more concentrated near the surface. These similar structures with respect to $\tau$ and space lead to good correlations between anomalous variables in the control and warmed climates (panel c), although $w'$ is somewhat noisy. Lastly, the anomalous variables as a function of time tend to stay within 30\% of their reference value (d), though there is some movement.

We more directly assess the importance of ENSO structure changes by evaluating Eq. \ref{Eq:MethodsEnergyBalancePrediction} using different ENSO composites and seeing how the results differ (Extended Data Fig. \ref{Fig:ED_WarmedComposite}). Using composites from the control or warmed state gives very similar time series, and changing one anomaly variable at a time does not qualitatively change the result. The exception is $w'$, for which the warmed version does not allow a period of enhanced variability. Comparing the control and warmed structure results directly has an $R^2$ value of 0.98 (panel b), showing that overall, changes to ENSO structure with warming play only a minor role in setting its amplitude, validating our use of a constant El Ni\~no structure over time.

\subsection*{GMST Prediction from an Energy Balance Model}

% Fig. \ref{Fig:LagLinearDerivationFigure} and Extended Data Tbl. 3 display predictions for ENSO variability without running any GCM realizations. These are calculated by using a FaIR three-box energy balance model to forecast GMST from an emissions scenario, then our lag-linear model to predict ENSO variability. The details of the GMST predictors are discussed here.

To approximate GMST in CMIP models we use the FaIR simple climate model \citep{leach2021fairv2} based on the energy balance model in \citet{cummins2020optimal}. Most model parameters are taken directly from the fits distributed with FaIR (\url{https://raw.githubusercontent.com/OMS-NetZero/FAIR/master/tests/test_data/4xCO2_cummins_ebm3.csv}), with the exception of a conversion from global mean surface air temperature to GMST, which is calculated via a linear fit for each model (calculated from a single SSP, as with the lag-linear model parameters). The GMST is then smoothed and compared to the 1960-1980 values to calculate $\Delta \overline{\mathrm{GMST}}$.

For the MITgcm, we use a single box energy balance:

\begin{equation}
    \frac{d}{dt} \Delta \overline{\mathrm{GMST}} = \frac{1}{\tau} \left[ \Delta T_{\infty} \mathrm{min}\left(\frac{t}{\Delta t},1\right) - \Delta\overline{\mathrm{GMST}}\right]
\end{equation}
so that the temperature approaches $\Delta T_{\infty}$ with the shape $\left(1-\exp(-t/\tau)\right)$ and the timescale $\tau$ is fit to the $\Delta t=50$ yr simulation.

% \section*{Declarations}



\backmatter

\bmhead{Acknowledgements}
We would like to thank the University of Chicago for research support and NCAR for the use of data stored on the Derecho computer system. D. Y. is supported by the NSF CAREER award and the Packard Fellowship. MIROC6 large ensemble data was provided by Michiya Hayashi. Kaustubh Thirumalai provided the CESM1 steady-state simulations. 

\clearpage

\bibliography{sn-bibliography}% common bib file

\section*{Extended Data Legends}

% --- Extended Data Figure Legends (Text Only) ---
% Note: Do not include the actual \includegraphics command here.
% Nature requires a list of legends.
\vspace{1em}
\extfiglegend{Fig:ED_UnsmoothedENSO}{Fig. 1 without smoothing}{As in Fig. 1, but values of $\Delta \mathrm{M}_{\mathrm{ENSO}}$ are not smoothed by a 20 year moving mean. Values of $R^2$ are lower due to more noise and unrelated internal variability. The CMIP ensemble has an $R^2$ of 0.89, CanESM5 SSP3-7.0 has an $R^2$ of 0.97 while all CanESM5 scenarios have an $R^2$ of 0.93, and MITgcm $\Delta t=50$\,yr has an $R^2$ of 0.99 while all MITgcm scenarios have an $R^2$ of 0.96.}

\vspace{1em}
\extfiglegend{Fig:ED_ElNinoVsLaNina}{Introduction to the simulations used in this study, and the differing warming rates of El Ni\~no and La Ni\~na events}{Panel a shows 20-year moving minima and maxima East Pacific SSTs under warming in the full climate model simulations (left) and the idealized MITgcm simulations (right). Panel b shows how the moving minima and maxima warming relate to changes in the moving mean temperature of the tropical ocean surface.}

\vspace{1em}
\extfiglegend{Fig:ED_SSPsLagLinear}{Applying the lag-linear predictor to a range of models and warming scenarios}{Panels a-e show time series of simulated ENSO variability (solid) and predicted ENSO variability from the lag-linear model (dashed). Color corresponds to warming scenario (i.e., $\Delta t$ value in panel a or SSP in the CMIP models), and the uncertainty corresponds to an n=100 bootstrap range. The lag-linear model fits very well in the MITgcm, CanESM5, and EC-Earth3. It fits slightly less well in CESM2 SSP2 and 5 due to the relatively few realizations available. It does not fit as well in MIROC6, possibly due to non-linearities present in that model but not represented by the lag-linear predictor. Panel f shows a scatterplot of all simulations used and the average R$^2$ value in each model -- all $R^2$ values are shown in Extended Data Tbl. 2.}

\vspace{1em}
\extfiglegend{Fig:ED_AbruptLagLinear}{Applying the lag-linear model to abrupt warming scenarios}{Panels a-c show time series of ENSO variability from three selected models under an abrupt quadrupling of CO$_2$ and the prediction of the lag-linear model. CanESM5 is chosen because it is used in Fig. 1 and Fig. 4, ACCESS-CM2 is chosen as a representative high $R^2$ model, and AWI-CM-1-1-MR is chosen as a representative low $R^2$ model. Panel d shows the ensemble mean (purple) and each individual model in gray. All models are displayed as a scatter of predicted vs. simulated ENSO variability in panel e, with the highlighted models and ensemble mean in color. Finally, panel f shows the relationship between the $R^2$ from the lag-linear fit and the ratio of forced to internal variability, demonstrating that the models with low $R^2$ tend to be dominated by internal variability. The internal variability of ENSO strength is calculated as the standard deviation of $\Delta \mathrm{M}_{\mathrm{ENSO}}$ in the pre-industrial climate, while the forced variability is calculated as the same quantity in the 80 years after the forcing begins. All $R^2$ and fit values are displayed in Extended Data Table 2. Uncertainty bands for individual models correspond to the size of the 2.5\%-97.5\% percentile range of that model's piControl simulation; for the ensemble mean it is calculated by an N=100 bootstrap resampling across models.}

\vspace{1em}
\extfiglegend{Fig:ED_Paleo}{Applying the lag-linear model to steady-state ENSO variability}{Panel a shows ENSO variability in a range of quasi-steady-state MITgcm simulations (averaged over 100 years at least 150 years after an abrupt forcing was applied), and the prediction from the lag-linear model with parameters fit to the $\Delta t=50$ yr simulations. Uncertainty bars show one standard deviation estimated by drawing 1000 bootstrap samples of 50 years selected with replacement. The lag-linear model accurately predicts ENSO strength in warmer climates, but fails in colder climates (not shown). Panel b shows ENSO variability in CESM1 simulations as in \citet{thirumalai_future_2024}, labeled with the time period they are meant to represent. Aside from GMST, orbital parameters and sea level (and therefore land-sea masks) are also changing over time. The dashed line represents the line of best fit through the five most recent scenarios, constrained to pass through the origin. Error bars are calculated as the 2.5\%-97.5\% range of an N=100 bootstrap resampling.}

\vspace{1em}
\extfiglegend{Fig:ED_Analytic}{Solving for ENSO variability analytically}{Panel a shows the MITgcm $\Delta t=50$ yr and CMIP abrupt quadrupling ensemble means decomposed into transient and equilibrium components. Panel b.i shows the simulated transient and full peak timing as a function of predicted transient component peak time, while b.ii shows the same for amplitude. Panel c shows the dependence of the non-dimensional peak amplitude (color) and time (contours) as a function of two relevant non-dimensional parameters. Details of the calculations are in Supplementary Text 4.}

\vspace{1em}
\extfiglegend{Fig:ED_WarmedComposite}{Understanding how the constant ENSO structure assumption affects the results}{Panel a shows the same time series as Fig.~2c but using the control or warmed structure or a mixture (i.e., most variables from the reference composite and one variable from the warmed composite), while panel b shows the relationship between the control and warmed ENSO structure predictions.}

\clearpage

\vspace{1em}
\noindent\textbf{Extended Data Table 1 $\mid$ Coefficients and fit metrics for the first linear model.}
The first section shows the slopes between mean quantities ($\langle \Delta \bar{T} \rangle$ or $\langle \Delta \bar{\Gamma} \rangle$) and the energy budget terms, as well as the goodness of fit. The second section shows the corresponding $A$ and $B$ values predicted from the energy budget compared to the empirically fit parameters. The corresponding time series are shown in Fig. S5.

% Generated by ConcatenateTableComponents on 15-Jun-2026 12:44:01
% Source: Table Components/Extended_Table_1/  (3 chunk(s))
% --- Heading + grouped energy budget term slopes (zero-intercept) ---
\begin{tabular}{llrr}
\toprule
Quantity & Predictor & Value & $R^2$ \\
\midrule
\multicolumn{4}{l}{\textit{Energy Budget Term}} \\
$\bar{\boldsymbol{u}}$ & $T$ & -0.072 & \textcolor[HTML]{5AAE61}{0.95} \\
$\boldsymbol{\nabla}_h \bar{T}$ & $T$ & -0.064 & \textcolor[HTML]{5AAE61}{0.94} \\
$\bar{\alpha}$ & $T$ & -0.055 & \textcolor[HTML]{1B7837}{0.97} \\
$\partial_z \bar{T}$ & $\Gamma$ & 9.084 & \textcolor[HTML]{2C7FB8}{0.86} \\
% --- First linear model coefficients ---
\midrule
\multicolumn{4}{l}{\textit{Linear Model}} \\
$A$ energy budget (1/K) & $T$ & -0.191 & \multirow{2}{*}{\textcolor[HTML]{5AAE61}{0.95}} \\
$B$ energy budget (m/K) & $\Gamma$ & 9.084 &        \\
$A$ fit (1/K)           & $T$ & -0.151 & \multirow{2}{*}{\textcolor[HTML]{1B7837}{0.95}} \\
$B$ fit (m/K)           & $\Gamma$ & 8.486 &        \\
% --- Footer ---
\bottomrule
\end{tabular}

\clearpage

\vspace{1em}
\noindent\textbf{Extended Data Table 2 $\mid$ Fit parameters for the lag-linear model.}
This table displays all lag-linear fits, calculated as discussed in the main text and methods section. Bold rows indicate the lag-linear model was fit to that scenario. Asterisks indicate that the model has a forced response to internal variability ratio less than 1.5 (Extended Data Fig. 4f). The fit parameter $t_{\mathrm{lag}}$ is constrained to be between 5 and 45 years and no more than half the length of the time series.

% Generated by ConcatenateTableComponents on 15-Jun-2026 12:44:02
% Source: Table Components/Extended_Table_2/  (3 chunk(s))
% --- Longtable header + MITgcm section (GlobalMean predictor) ---
\begin{longtable}{lrrrr}
\toprule
Scenario (Realizations) & $\alpha$ (1/K) & $\beta$ (1/K) & $t_{\mathrm{lag}}$ (yr) & $R^2$ \\
\midrule
\endhead
\rowcolor{gray!25} \multicolumn{5}{l}{\large\textbf{MITgcm}} \\
$\Delta t = 10\,\mathrm{yr}$ (25) & 0.028 & -0.061 & 34.6 & \textcolor[HTML]{1B7837}{0.98} \\
\textbf{$\boldsymbol{\Delta t = 50\,\mathrm{yr}}$ \textbf{(25)}} & \textbf{0.028} & \textbf{-0.061} & \textbf{34.6} & \textbf{\textcolor[HTML]{1B7837}{1.00}} \\
$\Delta t = 100\,\mathrm{yr}$ (25) & 0.028 & -0.061 & 34.6 & \textcolor[HTML]{1B7837}{0.98} \\
$\Delta t = 150\,\mathrm{yr}$ (25) & 0.028 & -0.061 & 34.6 & \textcolor[HTML]{5AAE61}{0.91} \\
\multicolumn{4}{c}{MITgcm Average R\(^2\)\quad \textcolor[HTML]{1B7837}{0.97}} & \\
\textit{Bootstrap CI} & [0.022, 0.034] & [-0.068, -0.055] & [31.6, 37.8] & [0.98, 1.00] \\
% --- SSP ensemble per-scenario rows (GlobalMean predictor) ---
\midrule
\rowcolor{gray!25} \multicolumn{5}{l}{\large\textbf{SSP Ensembles}} \\
CanESM5 SSP1-2.6 (50) & 0.227 & -0.205 & 16.6 & \textcolor[HTML]{1B7837}{0.95} \\
CanESM5 SSP2-4.5 (50) & 0.227 & -0.205 & 16.6 & \textcolor[HTML]{5AAE61}{0.95} \\
\textbf{CanESM5 SSP3-7.0 (50)} & \textbf{0.227} & \textbf{-0.205} & \textbf{16.6} & \textbf{\textcolor[HTML]{1B7837}{0.99}} \\
CanESM5 SSP5-8.5 (50) & 0.227 & -0.205 & 16.6 & \textcolor[HTML]{1B7837}{0.99} \\
\multicolumn{4}{c}{CanESM5 Average R\(^2\)\quad \textcolor[HTML]{1B7837}{0.97}} & \\
\textit{Bootstrap CI} & [0.145, 0.426] & [-0.399, -0.122] & [6.7, 28.8] & [0.95, 0.99] \\
\midrule
CESM2 SSP2-4.5 (16) & 0.247 & -0.307 & 27.2 & \textcolor[HTML]{7B5BA1}{0.77} \\
\textbf{CESM2 SSP3-7.0 (100)} & \textbf{0.247} & \textbf{-0.307} & \textbf{27.2} & \textbf{\textcolor[HTML]{5AAE61}{0.95}} \\
CESM2 SSP5-8.5 (15) & 0.247 & -0.307 & 27.2 & \textcolor[HTML]{7B5BA1}{0.74} \\
\multicolumn{4}{c}{CESM2 Average R\(^2\)\quad \textcolor[HTML]{2C7FB8}{0.82}} & \\
\textit{Bootstrap CI} & [0.163, 0.509] & [-0.553, -0.242] & [12.0, 45.0] & [0.77, 0.93] \\
\midrule
EC-Earth3 SSP1-2.6 (42) & 0.307 & -0.114 & 45.0 & \textcolor[HTML]{2C7FB8}{0.87} \\
\textbf{EC-Earth3 SSP2-4.5 (42)} & \textbf{0.307} & \textbf{-0.114} & \textbf{45.0} & \textbf{\textcolor[HTML]{1B7837}{0.96}} \\
EC-Earth3 SSP3-7.0 (42) & 0.307 & -0.114 & 45.0 & \textcolor[HTML]{1B7837}{0.97} \\
EC-Earth3 SSP5-8.5 (42) & 0.307 & -0.114 & 45.0 & \textcolor[HTML]{1B7837}{0.98} \\
\multicolumn{4}{c}{EC-Earth3 Average R\(^2\)\quad \textcolor[HTML]{5AAE61}{0.94}} & \\
\textit{Bootstrap CI} & [0.270, 0.534] & [-0.335, -0.041] & [12.2, 45.0] & [0.95, 0.99] \\
\midrule
MIROC6 SSP1-2.6 (50) & 0.663 & -0.823 & 30.0 & \textcolor[HTML]{555555}{0.40} \\
MIROC6 SSP2-4.5 (50) & 0.663 & -0.823 & 30.0 & \textcolor[HTML]{555555}{0.70} \\
\textbf{MIROC6 SSP3-7.0 (50)} & \textbf{0.663} & \textbf{-0.823} & \textbf{30.0} & \textbf{\textcolor[HTML]{1B7837}{0.99}} \\
MIROC6 SSP5-8.5 (50) & 0.663 & -0.823 & 30.0 & \textcolor[HTML]{2C7FB8}{0.87} \\
\multicolumn{4}{c}{MIROC6 Average R\(^2\)\quad \textcolor[HTML]{7B5BA1}{0.74}} & \\
\textit{Bootstrap CI} & [0.486, 0.768] & [-0.977, -0.664] & [27.5, 44.8] & [0.91, 0.98] \\
\midrule
\multicolumn{4}{c}{SSP Ensembles Average R\(^2\)\quad \textcolor[HTML]{2C7FB8}{0.87}} & \\
% --- Abrupt-4xCO2 rows + table footer (GlobalMean predictor) ---
\midrule
\rowcolor{gray!25} \multicolumn{5}{l}{\large\textbf{Abrupt-4xCO2 ensemble}} \\
Ensemble mean (46) & 0.261 & -0.216 & 17.9 & \textcolor[HTML]{1B7837}{0.98} \\
ACCESS-CM2 & 0.259 & -0.268 & 20.9 & \textcolor[HTML]{1B7837}{0.99} \\
ACCESS-ESM1-5 & 0.266 & -0.256 & 20.4 & \textcolor[HTML]{1B7837}{0.96} \\
AWI-CM-1-1-MR$^{*}$ & 0.051 & 0.061 & 40.0 & \textcolor[HTML]{555555}{0.55} \\
BCC-CSM2-MR & 0.162 & -0.274 & 13.5 & \textcolor[HTML]{1B7837}{0.98} \\
BCC-ESM1 & 0.166 & -0.251 & 18.9 & \textcolor[HTML]{1B7837}{0.99} \\
CAMS-CSM1-0$^{*}$ & 0.005 & 0.030 & 40.0 & \textcolor[HTML]{5AAE61}{0.91} \\
CESM2 & 0.370 & -0.376 & 12.5 & \textcolor[HTML]{5AAE61}{0.92} \\
CESM2-FV2$^{*}$ & 0.055 & -0.082 & 40.0 & \textcolor[HTML]{1B7837}{0.98} \\
CESM2-WACCM & 0.124 & -0.134 & 40.0 & \textcolor[HTML]{1B7837}{0.98} \\
CESM2-WACCM-FV2$^{*}$ & 0.049 & -0.112 & 36.3 & \textcolor[HTML]{1B7837}{1.00} \\
CIESM & 0.219 & -0.252 & 15.5 & \textcolor[HTML]{1B7837}{0.95} \\
CMCC-CM2-SR5$^{*}$ & 0.190 & 0.037 & 5.0 & \textcolor[HTML]{2C7FB8}{0.89} \\
CMCC-ESM2$^{*}$ & 0.152 & -0.016 & 5.0 & \textcolor[HTML]{7B5BA1}{0.71} \\
CNRM-CM6-1 & 0.225 & -0.204 & 20.7 & \textcolor[HTML]{1B7837}{0.97} \\
CNRM-ESM2-1$^{*}$ & 0.189 & -0.138 & 15.9 & \textcolor[HTML]{7B5BA1}{0.79} \\
CanESM5 & 0.464 & -0.469 & 16.3 & \textcolor[HTML]{1B7837}{0.96} \\
E3SM-1-0$^{*}$ & 0.222 & -0.161 & 19.9 & \textcolor[HTML]{1B7837}{0.98} \\
EC-Earth3 & 0.313 & -0.198 & 20.1 & \textcolor[HTML]{2C7FB8}{0.88} \\
EC-Earth3-AerChem & 0.665 & -0.458 & 5.0 & \textcolor[HTML]{2C7FB8}{0.80} \\
EC-Earth3-CC$^{*}$ & 0.250 & -0.137 & 40.0 & \textcolor[HTML]{1B7837}{0.99} \\
EC-Earth3-Veg$^{*}$ & 0.322 & -0.207 & 16.9 & \textcolor[HTML]{5AAE61}{0.95} \\
FGOALS-f3-L$^{*}$ & 0.290 & -0.249 & 17.8 & \textcolor[HTML]{2C7FB8}{0.87} \\
FIO-ESM-2-0 & 0.433 & -0.330 & 20.2 & \textcolor[HTML]{1B7837}{0.98} \\
GFDL-CM4 & 0.326 & -0.331 & 21.2 & \textcolor[HTML]{1B7837}{0.98} \\
GFDL-ESM4$^{*}$ & 0.249 & -0.238 & 16.5 & \textcolor[HTML]{5AAE61}{0.93} \\
GISS-E2-1-G$^{*}$ & 0.088 & -0.099 & 17.3 & \textcolor[HTML]{2C7FB8}{0.86} \\
GISS-E2-1-H$^{*}$ & 0.506 & -0.260 & 9.2 & \textcolor[HTML]{5AAE61}{0.91} \\
GISS-E2-2-G$^{*}$ & 0.157 & 0.118 & 38.0 & \textcolor[HTML]{5AAE61}{0.94} \\
GISS-E2-2-H & 0.104 & 0.257 & 28.9 & \textcolor[HTML]{7B5BA1}{0.80} \\
ICON-ESM-LR & 0.129 & -0.180 & 31.5 & \textcolor[HTML]{1B7837}{0.98} \\
INM-CM4-8 & 0.440 & -0.515 & 18.2 & \textcolor[HTML]{1B7837}{0.99} \\
INM-CM5-0 & 0.421 & -0.437 & 17.0 & \textcolor[HTML]{1B7837}{0.96} \\
IPSL-CM5A2-INCA & 0.268 & -0.344 & 17.2 & \textcolor[HTML]{1B7837}{1.00} \\
IPSL-CM6A-LR & 0.212 & -0.215 & 12.2 & \textcolor[HTML]{2C7FB8}{0.87} \\
MCM-UA-1-0 & 0.498 & -0.431 & 13.1 & \textcolor[HTML]{2C7FB8}{0.81} \\
MIROC-ES2H$^{*}$ & 0.366 & -0.183 & 22.4 & \textcolor[HTML]{1B7837}{0.97} \\
MIROC-ES2L$^{*}$ & 0.315 & -0.233 & 13.6 & \textcolor[HTML]{5AAE61}{0.92} \\
MIROC6$^{*}$ & 0.403 & -0.269 & 23.4 & \textcolor[HTML]{1B7837}{0.98} \\
MPI-ESM-1-2-HAM & 0.235 & -0.321 & 19.4 & \textcolor[HTML]{1B7837}{0.95} \\
MPI-ESM1-2-HR$^{*}$ & 0.028 & -0.003 & 39.3 & \textcolor[HTML]{2C7FB8}{0.88} \\
MPI-ESM1-2-LR$^{*}$ & 0.214 & -0.194 & 13.4 & \textcolor[HTML]{7B5BA1}{0.72} \\
MRI-ESM2-0$^{*}$ & 0.086 & 0.052 & 40.0 & \textcolor[HTML]{2C7FB8}{0.88} \\
NESM3 & 0.470 & -0.438 & 11.9 & \textcolor[HTML]{2C7FB8}{0.85} \\
NorCPM1$^{*}$ & -0.127 & -0.057 & 40.0 & \textcolor[HTML]{1B7837}{0.98} \\
NorESM2-LM$^{*}$ & 0.355 & -0.172 & 5.0 & \textcolor[HTML]{555555}{0.59} \\
UKESM1-0-LL & 0.083 & -0.157 & 40.0 & \textcolor[HTML]{5AAE61}{0.92} \\
\multicolumn{4}{c}{Abrupt Average R\(^2\)\quad \textcolor[HTML]{5AAE61}{0.91}} & \\
\textit{Bootstrap CI} & [0.233, 0.287] & [-0.243, -0.188] & [16.7, 19.2] & [0.96, 0.99] \\
\bottomrule
\end{longtable}

\clearpage

\vspace{1em}
\noindent\textbf{Extended Data Table 3 $\mid$ Using few realizations to predict ensemble mean ENSO variability.}
This table tests the predictive skill of the lag-linear model compared with a GCM ensemble when using different numbers of realizations. For each model and scenario, the columns display: 1. the correlation between full-ensemble ENSO variability and that predicted by an Energy Balance Model combined with the lag-linear predictor (see methods section), 2. the correlation of a single realization with full-ensemble ENSO variability if using the simulated ENSO variability (left) or that predicted by the lag-linear model (right), 3. the number of realizations required to achieve an $R^2$ of 0.9 through direct simulation of ENSO variability, and 4. the number of realizations required for the diagnosed ENSO variability from the simulations to match that predicted by the lag-linear model. Bold rows indicate the lag-linear model was fit to that scenario.

% --- Extended Data Table 3 ---
% Realizations needed for simulated R^2 to reach 0.9 and to cross the
% lag-linear prediction; plus the FaIR-emulator zero-realization R^2
% and simulated/predicted R^2 at N=1.
% Bootstrap range: 1 to Inf. Training scenario row is bolded.
% Requires: \usepackage{xcolor, colortbl, booktabs, longtable}.
\begin{longtable}{l@{\hspace{2pt}}ccccc}
\toprule
 & $R^2$ at $N{=}0$ & \multicolumn{2}{c}{$R^2$ at $N{=}1$} & & \\
\cmidrule(lr){3-4}
 & (EBM) & Simulated & Lag-linear & $N(R^2_\mathrm{sim} = 0.9)$ & $N(R^2_\mathrm{sim} = R^2_\mathrm{pred})$ \\
\midrule
\endhead
\rowcolor{gray!25} \multicolumn{6}{l}{\textbf{MITgcm}} \\
$\Delta t = 10$ yr & \textcolor[HTML]{1B7837}{0.96} & \textcolor[HTML]{5AAE61}{0.91} & \textcolor[HTML]{1B7837}{0.98} & 1 & 5 \\
{\boldmath\textbf{$\Delta t = 50$ yr}} & \textcolor[HTML]{5AAE61}{0.95} & \textcolor[HTML]{2C7FB8}{0.89} & \textcolor[HTML]{1B7837}{1.00} & \textbf{2} & \textbf{14} \\
$\Delta t = 100$ yr & \textcolor[HTML]{1B7837}{0.96} & \textcolor[HTML]{2C7FB8}{0.85} & \textcolor[HTML]{1B7837}{0.98} & 2 & 6 \\
$\Delta t = 150$ yr & \textcolor[HTML]{5AAE61}{0.94} & \textcolor[HTML]{7B5BA1}{0.78} & \textcolor[HTML]{5AAE61}{0.91} & 3 & 3 \\
\midrule
\rowcolor{gray!25} \multicolumn{6}{l}{\textbf{SSP Ensembles}} \\
CanESM5 SSP1-2.6 & \textcolor[HTML]{2C7FB8}{0.86} & \textcolor[HTML]{555555}{0.28} & \textcolor[HTML]{2C7FB8}{0.86} & 20 & 17 \\
CanESM5 SSP2-4.5 & \textcolor[HTML]{1B7837}{0.97} & \textcolor[HTML]{555555}{0.38} & \textcolor[HTML]{1B7837}{0.96} & 10 & 23 \\
{\boldmath\textbf{CanESM5 SSP3-7.0}} & \textcolor[HTML]{1B7837}{0.99} & \textcolor[HTML]{555555}{0.59} & \textcolor[HTML]{1B7837}{0.99} & \textbf{6} & \textbf{34} \\
CanESM5 SSP5-8.5 & \textcolor[HTML]{1B7837}{0.99} & \textcolor[HTML]{555555}{0.69} & \textcolor[HTML]{1B7837}{0.99} & 4 & 22 \\
{\boldmath\textbf{CESM2 SSP3-7.0}} & \textcolor[HTML]{2C7FB8}{0.89} & \textcolor[HTML]{555555}{0.32} & \textcolor[HTML]{5AAE61}{0.90} & \textbf{19} & \textbf{30} \\
EC-Earth3 SSP1-2.6 & \textcolor[HTML]{2C7FB8}{0.89} & \textcolor[HTML]{555555}{0.49} & \textcolor[HTML]{2C7FB8}{0.87} & 8 & 7 \\
{\boldmath\textbf{EC-Earth3 SSP2-4.5}} & \textcolor[HTML]{1B7837}{0.97} & \textcolor[HTML]{555555}{0.65} & \textcolor[HTML]{1B7837}{0.95} & \textbf{6} & \textbf{15} \\
EC-Earth3 SSP3-7.0 & \textcolor[HTML]{1B7837}{0.97} & \textcolor[HTML]{7B5BA1}{0.72} & \textcolor[HTML]{1B7837}{0.96} & 4 & 12 \\
EC-Earth3 SSP5-8.5 & \textcolor[HTML]{1B7837}{0.98} & \textcolor[HTML]{7B5BA1}{0.79} & \textcolor[HTML]{1B7837}{0.97} & 3 & 11 \\
MIROC6 SSP1-2.6 & \textcolor[HTML]{555555}{0.35} & \textcolor[HTML]{555555}{0.39} & \textcolor[HTML]{555555}{0.40} & 12 & 2 \\
MIROC6 SSP2-4.5 & \textcolor[HTML]{555555}{0.47} & \textcolor[HTML]{555555}{0.52} & \textcolor[HTML]{555555}{0.64} & 9 & 3 \\
{\boldmath\textbf{MIROC6 SSP3-7.0}} & \textcolor[HTML]{2C7FB8}{0.87} & \textcolor[HTML]{555555}{0.61} & \textcolor[HTML]{1B7837}{0.95} & \textbf{6} & \textbf{26} \\
MIROC6 SSP5-8.5 & \textcolor[HTML]{1B7837}{0.97} & \textcolor[HTML]{555555}{0.45} & \textcolor[HTML]{2C7FB8}{0.84} & 9 & 7 \\
\bottomrule
\end{longtable}

\end{document}

%\linenumbers

\title[Article Title]{Supplemental Information: The Rise and Fall of ENSO in a Warming World: Insights from a Lag-Linear Model}

\author*[1,2,3]{\fnm{P.J.} \sur{Tuckman}}\email{pjt5@stanford.edu}

\author*[1,2]{\fnm{Da} \sur{Yang}}\email{dayang@stanford.edu}

\affil[1]{\orgdiv{Department of Geophysical Sciences}, \orgname{University of Chicago}, \orgaddress{\street{5734 S Ellis Ave}, \city{Chicago}, \state{IL}, \postcode{60637}, \country{United States of America}}}

\affil[2]{\orgdiv{Department of Geophysics}, \orgname{Stanford University}, \orgaddress{\street{397 Panama Mall}, \city{Stanford}, \state{CA}, \postcode{94305},  \country{United States of America}}}

\affil[3]{\orgname{University Corporation for Atmospheric Research}, \orgaddress{\street{3090 Center Green Dr}, \city{Boulder}, \state{CO}, \postcode{80301},  \country{United States of America}}}

% \affil[2]{\orgdiv{Department of Earth, Atmospheric, and Planetary Sciences}, \orgname{Massachusetts Institute of Technology}, \orgaddress{\street{21 Ames Street}, \city{Cambridge}, \postcode{02139}, \state{MA}, \country{United States of America}}}

\renewcommand{\thefigure}{S.\arabic{figure}}
\setcounter{figure}{0}

\maketitle

% --- Table of Contents ---
\vspace{1em}

\noindent
Supplementary Text 1: Surface Fluxes and Warming \dotfill Page 2 \\
Supplementary Text 2: Predicting Linear Model Coefficients \dotfill Page 4 \\
Supplementary Text 3: Connecting the Linear Models \dotfill Page 7 \\
Supplementary Text 4: Analytic Solutions to the Lag-linear Model \dotfill Page 9 \\
Supplementary Figures 1 to 11 \dotfill Page 13
%\tableofcontents

\clearpage

\section{Surface Fluxes and Warming}{\label{sec:SurfaceFluxes}}

Here, the weakening of ENSO events due to strengthened surface flux damping is discussed in detail. We find that this effect is controlled by latent heat fluxes and driven by the increase in saturation specific humidity with warming. 

The sum of sub-grid-scale processes (i.e., radiation and turbulent fluxes) is a damping effect in all climates, meaning it is negatively correlated with anomalous surface temperatures (Fig. S10). When the mean temperature is larger, i.e., in a warmer climate, the slope of this relationship increases, making surface fluxes a stronger damping effect. Anomalous latent heat flux (LHF$'$, scaled by the heat capacity of the mixed layer so it has units of temperature tendency) shows these same features (Fig. S10) -- it is a damping effect and the relevant slope steepens with warming. 

The strength of total surface flux damping, represented by $\bar{\alpha}_{\mathrm{SF}}$ in the main text, is shown as a function of time in Fig. S10c and, as expected, it becomes more negative with warming. To decompose $\overline{\alpha}_\mathrm{SF}$, we define an $\overline{\alpha}$ for each variable such that $\overline{\alpha}_x$ is the slope of the relationship between $x'$ and SST$'$ for positive values of SST$'$ within 10 years on either side of time $t$. The LHF and total $\overline{\alpha}$ values are similar, and the increase with warming of the total $\bar{\alpha}_{\mathrm{SF}}$ is controlled by that of LHF. Therefore, the strengthening of surface flux damping with warming is associated with LHF, i.e. evaporation from the ocean surface into the boundary layer. 

To understand why $\overline{\alpha}_{\mathrm{LHF}}$ increases with warming, we consider the bulk formula for evaporation in climate models:

\begin{equation}
\label{eq:BulkFormulaEvap}
    \mathrm{LHF}=C_d |\vec{u}| q^*(1-\mathrm{RH}) = C q^*(1-\mathrm{RH})
\end{equation}
where $C_d$ is a constant coefficient, $|\vec{u}|$ is the wind speed, and $q^*(1-\mathrm{RH})$ is the saturation deficit, or the difference between the saturation specific humidity of the SST and the specific humidity of the lowest atmospheric level. The second expression consolidates the two variables which do not change significantly with warming into a constant $C$. We wish to study anomalous fluxes:

\begin{equation}
    \mathrm{LHF}' \approx C' \overline{q^*} \overline{(1-\mathrm{RH})} + \overline{C} q'^* \overline{(1-\mathrm{RH})} - \overline{C} \overline{q^*} \mathrm{RH}'
\end{equation}
where non-linear terms are ignored. Calculating $\overline{\alpha}$ values for each anomalous variable, we find:

\begin{equation}
    \label{eq:LHFAlphaComponents}
    \overline{\alpha}_{\mathrm{LHF}} \approx \overline{\alpha}_C \overline{q^*} \overline{(1-\mathrm{RH})} + \overline{C} \overline{\alpha}_{q^*} \overline{(1-\mathrm{RH})} - \overline{C} \overline{q^*} \overline{\alpha}_\mathrm{RH}.
\end{equation}

These three components and their sum are shown in Fig. S10. Their sum is almost identical to $\overline{\alpha}_{\mathrm{LHF}}$, indicating that the non-linear terms are indeed small. More importantly, we see that the $- \overline{C} \overline{q^*} \overline{\alpha}_\mathrm{RH}$ term is causing most of the change in the overall surface flux damping. To understand why this is, we examine the impact of changing $\overline{q^*}$ or $\overline{\alpha}_\mathrm{RH}$ only, and see that changes in $\overline{q^*}$ are causing the overall decrease in $\overline{\alpha}_{\mathrm{LHF}}$. In fact, this term alone has a decrease of about $0.6$ 1/year between the reference and warmed climate, while the total $\overline{\alpha}_{\mathrm{SF}}$ decreases by 0.65 (panel c). In other words, the $\overline{\alpha}_\mathrm{RH}$ term is causing most of the change with warming, and the change within this term is due to increasing $\overline{q^*}$. Physically, the increase in saturation specific humidity with warming ($\overline{q^*}(\bar{T})$) causes a given anomalous relative humidity ($\mathrm{RH}'$) to be associated with a larger saturation deficit ($q^*(1-\mathrm{RH})$), and therefore more latent heat fluxes, in a warmer climate. Thus, the increase in saturation specific humidity caused by the Clausius-Clapeyron relationship leads to strengthened surface flux damping in a warmer climate. 

\clearpage

\section{Predicting Linear Model Coefficients from Energy Balance}{\label{sec:EmpiricalCoefficientsDerivation}}

We here use East Pacific energy balance and the MITgcm $\Delta t=50$ year simulations to calculate and interpret the coefficients of mean temperature and stratification in the linear model for ENSO amplitude. In other words, we begin with Eq. 3 and 4 from the main text then derive the expressions in Eq. 6. Our starting point, the energy balance expression for ENSO magnitude, is: 

\begin{equation*}
    \Delta \mathrm{M}_{\mathrm{ENSO}} (t) = \exp\!\left[\int_{\tau_0}^0 d\tau\,\left(-\frac{\langle \Delta \mathbf{\bar{u}}\!\cdot\!\nabla T' \rangle}{\langle T' \rangle}-\frac{\langle \mathbf{u'}_h \cdot \Delta \nabla_h \bar{T} \rangle}{\langle T' \rangle} -\frac{\langle w' \Delta \bar{\Gamma} \rangle}{\langle T' \rangle} -\Delta \bar{\alpha}_{\mathrm{SF}}\,\frac{\langle T' \rangle_{\mathrm{2D,surf}}}{\langle T' \rangle} \right)\right]-1.
\end{equation*}

The argument to the exponential is small, so we Taylor expand and keep only the linear term to find

\begin{equation*}
        \Delta \mathrm{M}_{\mathrm{ENSO}} (t) = \int_{\tau_0}^0 d\tau\,\left(-\frac{\langle \Delta \mathbf{\bar{u}}\!\cdot\!\nabla T' \rangle}{\langle T' \rangle}-\frac{\langle \mathbf{u'}_h \cdot \Delta \nabla_h \bar{T} \rangle}{\langle T' \rangle} -\frac{\langle w' \Delta \bar{\Gamma} \rangle}{\langle T' \rangle} -\Delta \bar{\alpha}_{\mathrm{SF}}\,\frac{\langle T' \rangle_{\mathrm{2D,surf}}}{\langle T' \rangle} \right).
\end{equation*}

Fig. 3a and b showed that each term in this expression is proportional to either changes in mean temperature or stratification. The somewhat non-linear relationship between stratification and its energy budget term is likely due to changes in the vertical structure of stratification -- the very shallow part of the ocean (e.g. the top 30m) warms earlier than the rest of the mixed layer, and the mean stratification does not fully capture this change. Extended Data Table 1 shows details of the linear relationships between mean quantities and the energy budget terms, displaying the proportionality constants and $R^2$ values. Given these proportionalities, we can define constants such that: 

\begin{equation}
        \Delta \mathrm{M}_{\mathrm{ENSO}} (t) = \left(A_{\bar{\mathbf{u}}} + A_{\nabla_h \bar{T}} + A_{\alpha}\right)\langle \Delta \bar{T} \rangle  +B\langle \Delta \bar{\Gamma} \rangle
\end{equation}
where:

\begin{equation}
    \begin{split}
        A_{\mathbf{\bar{u}}} &= \frac{1}{\langle \Delta \bar{T} \rangle} \int_{\tau_0}^0 d\tau \left( \frac{-\langle \Delta \bar{\mathbf{u}} \cdot \nabla T' \rangle }{\langle T' \rangle} \right) \\
        A_{\nabla_h \bar{T}} &= \frac{1}{\langle \Delta \bar{T} \rangle} \int_{\tau_0}^0 d\tau \left( \frac{-\langle \mathbf{u}_h' \cdot \Delta \nabla_h \bar{T} \rangle }{\langle T' \rangle} \right) \\
        A_{\bar{\alpha}_{\mathrm{SF}}} &= \frac{1}{\langle \Delta \bar{T} \rangle} \int_{\tau_0}^0 d\tau \left( -\Delta \bar{\alpha}_{\mathrm{SF}} \frac{\langle T' \rangle_{\mathrm{2D,surf}} }{\langle T' \rangle} \right) \\
        B &= \frac{1}{\langle \Delta \bar{\Gamma} \rangle} \int_{\tau_0}^0 d\tau \left( \frac{-\langle w' \Delta  \bar{\Gamma} \rangle }{\langle T' \rangle} \right). \\
    \end{split}
\end{equation}
To obtain this equation, we have simply multiplied each term by unity in the form of either $\langle \Delta \bar{T} \rangle/\langle \Delta \bar{T} \rangle$ or $\langle \Delta  \bar{\Gamma} \rangle/\langle \Delta \bar{\Gamma} \rangle$.

Intuitively, the $A$ terms are inverse temperature scales that represent the mean climate warming over which ENSO strength changes by a factor of $e$ (before Taylor expanding), and $B$ is similar but for stratification. Summing the $A$ terms, our final expression is:

\begin{equation}
     \Delta \mathrm{M}_{\mathrm{ENSO}} (t) =A\langle \Delta \bar{T} \rangle + B \langle \Delta \bar{\Gamma} \rangle
\end{equation}
where $A=A_{\mathbf{\bar{u}}}+ A_{\nabla_h \bar{T}} + A_{\bar{\alpha}_{\mathrm{SF}}}$.

Grouping the terms proportional to mean temperature, we can predict the linear model coefficients as:
\begin{equation}
    \begin{split}
        A &= \frac{1}{\langle \Delta \bar{T} \rangle} \left[ \int_{\tau_0}^0 d\tau \left( \frac{-\langle \Delta \bar{\mathbf{u}} \cdot \nabla T' \rangle}{\langle T' \rangle}  + \frac{-\langle \mathbf{u}_h' \cdot \Delta \nabla_h \bar{T} \rangle}{\langle T' \rangle} - \Delta \bar{\alpha}_{\mathrm{SF}} \frac{\langle T' \rangle_{\mathrm{2D,surf}}}{\langle T' \rangle} \right) \right] \\
        B &= \frac{1}{\langle \Delta \bar{\Gamma} \rangle} \int_{\tau_0}^0 d\tau \left( \frac{-\langle w' \Delta \bar{\Gamma} \rangle}{\langle T' \rangle} \right).
    \end{split}
\end{equation}
These expressions, calculated as the proportionality constant between the energy balance term and the mean quantity across all $\Delta t=50$ year data, give the values shown in Extended Data Tbl. 1, within 25\% ($A$) and 7\% ($B$) of the empirical best fits. Using these values instead of the empirical fit gives similar predictions for $\Delta \mathrm{M}_{\mathrm{ENSO}}$ and a very high $R^2$, shown in Fig. S5  and Extended Data Tbl. 1. To summarize, we can use energy balance to qualitatively and quantitatively show that changes to ENSO with warming are accurately described by only mean temperature and stratification, enabling a simple linear predictor to capture much of the future of ENSO variability.

These physically motivated expressions for the relevant coefficients also give insight into what controls ENSO variability. Using the stratification term as an example, we can move the mean stratification into the integral and multiply by $\langle w' \rangle/\langle w' \rangle$ to write:

\begin{equation}
    B = -\int_{\tau_0}^0 d\tau \frac{\langle w' \rangle}{\langle T' \rangle} \frac{\langle w'  \Delta \bar{\Gamma}\rangle}{\langle w' \rangle \langle \Delta \bar{\Gamma} \rangle}
\end{equation}

The first part of the integrand is an upwelling efficiency, measuring how much East Pacific upwelling changes per degree warming of an ENSO event (similar to a Bjerknes index coefficient). The second term is a correlation factor between $w'$ and $\Delta \bar{\Gamma}$, telling us that the more the two variables are aligned in space, the more effective changes in stratification are at modifying ENSO magnitude. Putting these together, we recover the intuitive and physically meaningful conclusion that the degree to which mean stratification changes ENSO magnitude (i.e., $B$), depends on 1. the normalized upwelling anomaly and 2. to what extent that upwelling is spatially aligned with the stratification changes. 

The same analysis can be applied to $A$, leading to:
\begin{equation}
    \begin{split}
        A_{\bar{\mathbf{u}}} &= -\frac{\langle \Delta\bar{\mathbf{u}}\rangle}{\langle \Delta\bar{T}\rangle}\cdot \int_{\tau_0}^{0} d\tau\, \frac{\langle \nabla T'\rangle}{\langle T'\rangle}\, \frac{\langle \Delta\bar{\mathbf{u}} \cdot \nabla T'\rangle}{\langle \Delta\bar{\mathbf{u}}\rangle \langle \nabla T'\rangle} \\
        A_{\nabla_h\bar{T}} &= -\frac{\langle \Delta\nabla_h\bar{T}\rangle}{\langle \Delta\bar{T}\rangle}\cdot \int_{\tau_0}^{0} d\tau\, \frac{\langle \mathbf{u}'_h\rangle}{\langle T'\rangle}\, \frac{\langle \mathbf{u}'_h \cdot \Delta\nabla_h\bar{T}\rangle}{\langle \mathbf{u}'_h\rangle\langle \Delta\nabla_h\bar{T}\rangle} \\
        A_{\bar{\alpha}_{\mathrm{SF}}} &= -\frac{\Delta\bar{\alpha}_{\mathrm{SF}}}{\langle \Delta\bar{T}\rangle} \int_{\tau_0}^{0} d\tau\, \frac{\langle T'\rangle_{\mathrm{2D,surf}}}{\langle T'\rangle}
    \end{split}
\end{equation}

Each advection term has an interpretation similar to the above, with the additional factor of how each mean variable changes per degree climate warming (e.g., $\langle\bar{\bf{u}}\rangle/\langle \bar{T}\rangle$). For surface flux damping, the relevant quantity is simply the change in $\bar{\alpha}$ per degree mean climate warming.

\clearpage

\section{Connecting the Linear Models}{\label{sec:LinearModelConnection}}

In this section, we quantitatively connect the two linear models for ENSO variability:

\begin{equation}
    \begin{split}
        \Delta \mathrm{M}_{\mathrm{ENSO}} &= A \langle \Delta \bar{T} \rangle + B \langle \Delta \bar{\Gamma} \rangle \\
        &= \alpha \Delta \overline{\mathrm{GMST}}(t) + \beta \Delta \overline{\mathrm{GMST}} (t-t_{\mathrm{lag}}) 
    \end{split}
\end{equation}
Relating these models requires linear relationships between the changes in relevant variables under warming; we represent these proportionalities with slopes $m_n$:
\begin{itemize}
    \item \textbf{$m_1$}: East Pacific mixed-layer temperature predicted by the average of East Pacific surface and subsurface temperature
    \item \textbf{$m_2$}: East Pacific subsurface temperature predicted by lagged subtropical surface temperature
    \item \textbf{$m_3$}: East Pacific surface temperature predicted by GMST
    \item \textbf{$m_4$}: Subtropical surface temperature predicted by GMST
\end{itemize}

Replacing stratification with the difference between East Pacific surface temperature $\langle \Delta \bar{T}\rangle_{\mathrm{2D,surf}}$ and East Pacific subsurface temperature $\langle \Delta \bar{T} \rangle_{\mathrm{2D,subsurf}}$, then assuming East Pacific mixed layer temperature $\langle \Delta \bar{T} \rangle$ is proportional to the average of East Pacific surface and subsurface temperature via slope $m_1$:

\begin{equation}
        \Delta \mathrm{M}_{\mathrm{ENSO}} = A m_1 \left[\langle \Delta \bar{T} \rangle_{\mathrm{2D,surf}} + \langle \Delta \bar{T} \rangle_{\mathrm{2D,subsurf}} \right]/2 + \frac{B}{\Delta z} \left[\langle \Delta \bar{T} \rangle_{\mathrm{2D,surf}} - \langle \Delta \bar{T} \rangle_{\mathrm{2D,subsurf}}\right].
\end{equation}

Next, we replace subsurface temperature with a constant $m_2$ multiplied by subtropical surface temperatures with a lag, and assume East Pacific and subtropical surface temperatures are both proportional to global mean surface temperature via slopes $m_3$ and $m_4$:

\begin{equation}
    \Delta \mathrm{M}_{\mathrm{ENSO}} = \left[\frac{A m_1}{2} + \frac{B}{\Delta z} \right] m_3  \Delta \overline{\mathrm{GMST}}(t)  + \left[\frac{Am_1}{2} - \frac{B}{\Delta z} \right] m_2 m_4 \Delta \overline{\mathrm{GMST}} (t-t_{\mathrm{lag}}) 
\end{equation}

Comparing these expressions to the lag-linear model:

\begin{equation}
    \begin{split}
        \alpha &= \left[\frac{A m_1}{2} + \frac{B}{\Delta z} \right] m_3 \\
        \beta &= \left[\frac{A m_1}{2} - \frac{B}{\Delta z} \right] m_2 m_4
    \end{split}
\end{equation}

Using the lines of best fit displayed in Fig. \ref{Fig:GMSTProportionality} and the fitted values of $A$ and $B$, we find that $\alpha=0.034$ 1/K and $\beta=-0.072$ 1/K, within 20-25\% of the fitted values. If we use energy budget predictions for $A$ and $B$, the predicted values are $\alpha=0.03$ 1/K and $\beta=-0.082$ 1/K, still within about 30\% of those displayed in Extended Data Table 2. The resulting predicted time series are shown in Fig. \ref{Fig:GMSTProportionality}c, and both versions roughly replicate the transient peak and long-term decline of ENSO variability. 

Additionally, these expressions suggest physical interpretations of the $\alpha$ and $\beta$ parameters. The response of ENSO to the surface is controlled by $\alpha$, given as a linear combination of $A$ (representing the response of the mixed-layer) and $B$ (representing the response of stratification). The response of ENSO to the subsurface is controlled by $\beta$, and $B$ therefore appears with a negative sign in that expression as stratification depends inversely on the temperature at depth. Overall, our physically motivated predictions of $\alpha$ and $\beta$, despite cumulative uncertainty from several proportionality constants, are close enough to reasonably predict ENSO variability, giving confidence in our expressions for $\alpha$ and $\beta$ in terms of physically meaningful parameters.

%We can also analyze inter-model differences in $\alpha$ and $\beta$ using these expressions. The difference between MITgcm and CESM in the first linear model is dominated by differing $B$ values, and since $B$ appears in the expressions for $\alpha$ and $\beta$, both parameters in the lag-linear model are significantly different. This suggests that the inter-model spread in representing how ENSO changes with warming is largely controlled by the Ekman feedback, as that is the physical process behind the value of $B$.

\clearpage

\section{Analytic Solutions to the Lag-linear Model}\label{sec:AnalyticSolution}

The lag-linear model predicts that ENSO variability depends only on global mean surface temperature over time and the parameters $\alpha$, $\beta$, and $t_\mathrm{lag}$. Here, we seek to understand the solution by relating ENSO strength over time, and especially its peak (defined to be at $t^{\mathrm{max}}$ and of amplitude $\Delta \mathrm{M}_{\mathrm{ENSO}}^{\mathrm{max}}$) to interpretable features of the temperature trajectory. 

We begin by separating the ENSO prediction into two components (Extended Data Fig. 6a): 

\begin{equation}
\Delta \mathrm{M}_{\mathrm{ENSO}}(t) \;=\; \underbrace{(\alpha+\beta)\,\Delta \overline{\mathrm{GMST}}(t)}_{\Delta \mathrm{M}_\mathrm{eq}(t)} \;+\; \underbrace{-\beta\bigl[\Delta \overline{\mathrm{GMST}}(t)-\Delta \overline{\mathrm{GMST}} (t-t_{\mathrm{lag}})\bigr]}_{\Delta \mathrm{M}_\mathrm{tr}(t)},
\label{eq:decomposition}
\end{equation}
where the first part is the ``evolving equilibrium solution," or the prediction if the system were in equilibrium at the current global mean surface temperature, and the second part is the ``transient solution," or the prediction if ENSO's steady-state response to warming were no change (i.e., $\alpha+\beta=0$).

If we consider the transient solution alone:

\begin{equation}
    \Delta \mathrm{M}_\mathrm{tr}(t)=-\beta\bigl[\Delta  \overline{\mathrm{GMST}} (t) -\Delta  \overline{\mathrm{GMST}}(t-t_{\mathrm{lag}})\bigr],
    \label{Eq:ST4TransientPart}
\end{equation}
some properties of its peak become immediately apparent. First, as $\beta$ is a constant, the largest ENSO variability will be when $\Delta  \overline{\mathrm{GMST}}(t)-\Delta  \overline{\mathrm{GMST}}(t-t_{\mathrm{lag}})$ is largest, i.e., at the end of the period of length $t_\mathrm{lag}$ when the most warming occurs. In an abrupt warming scenario, where temperatures rise fastest at the moment greenhouse gases are added, this will be at time $t_\mathrm{lag}$ after the warming begins. In a smooth scenario, the relevant temperature change is roughly proportional to a derivative evaluated at $t-t_{\mathrm{lag}}/2$, so if we refer to the time of fastest warming as $t_0$, then maximum ENSO variability occurs at $t_0+t_{\mathrm{lag}}/2$. 

Eq. \ref{Eq:ST4TransientPart} reveals that warming amplitude and timescale control peak ENSO variability. Peak variability occurs at the end of a fixed length period with the most warming, and its amplitude is proportional to the amount of warming in that time. This immediately suggests the importance of a warming amplitude ($\Delta T_{\infty}$) and timescale ($\tau_W$), the latter of which must be compared to $t_{\mathrm{lag}}$. Next, we must understand how the equilibrium component alters this predicted peak.

% To summarize, the transient component of ENSO variability peaks at the end of the period of length $t_\mathrm{lag}$ when the most warming occurs, and the peak's amplitude increases with warming amount and decreases with warming timescale. We now interpret the equilibrium component by studying how it will alter that peak. 

As we are studying a warming climate, we will assume $\Delta  \overline{\mathrm{GMST}}$ is positive and increasing for all values of $t$. Therefore, the signs of 

\begin{equation}
    \Delta \mathrm{M}_\mathrm{eq}(t)=(\alpha+\beta)\,\Delta  \overline{\mathrm{GMST}}(t)
\end{equation}
and its derivative are always set by the sign of $\alpha+\beta$, or the steady-state response of ENSO to increased temperatures. If ENSO is weaker in a warmer climate (i.e., $\alpha+\beta<$0), then $\Delta \mathrm{M}_\mathrm{eq}<0$ and $d\Delta \mathrm{M}_\mathrm{eq}/dt<0$, which will make ENSO's peak earlier and smaller than that predicted by the transient component alone. On the other hand, if ENSO is stronger in a warmer climate, ENSO's peak amplitude will be later and larger (Extended Data Fig. 6b). 

We can make further progress towards understanding the predictions of the lag-linear model by assuming tractable forms of the warming temperature trajectory; we now do this separately for the case of abrupt warming and realistic (though idealized) emissions scenarios. 

\subsection{Idealized Smooth Warming}

We assume a warming trajectory of the form:

\begin{equation}
    \Delta  \overline{\mathrm{GMST}}(t) = \frac{\Delta T_{\infty}}{2} \left[1+ \mathrm{erf} \left(\frac{t-t_0}{\tau_W} \right) \right]
    \label{eq:ERFTemperatureTrajectory}
\end{equation}
corresponding to a Gaussian derivative of temperature (and similar, but not identical, to a Gaussian shape of emissions). The relevant parameters are the final warming amount ($\Delta T_{\infty}$), a parameter representing the time of maximum warming $(t_0)$, and a warming timescale ($\tau_W$). This trajectory fits well to SSP simulations (Fig. S6). 

To find the maximum ENSO amplitude we set the time derivative of the lag-linear model to zero and combine with Eq. \ref{eq:ERFTemperatureTrajectory} to find:

\begin{equation}
    \alpha \exp{\left[-\left(\frac{t-t_0}{\tau_W}\right)^2\right]} = -\beta \exp{\left[-\left(\frac{t-t_\mathrm{lag}-t_0}{\tau_W}\right)^2\right]}
\end{equation}

Dividing by the exponential on the right side, then applying a natural logarithm to both sides:

\begin{comment}
\begin{equation}
    \begin{split}
        \alpha \exp\left[-\left(\frac{t-t_0}{\tau_W} \right)^2 + \left( \frac{t-t_{\mathrm{lag}}-t_0}{\tau_W}^2 \right) \right] &= -\beta \\
        \\
        -\left(\frac{t-t_0}{\tau_W} \right)^2 + \left( \frac{t-t_{\mathrm{lag}}-t_0}{\tau_W} \right)^2 &= \ln{\left(\frac{-\beta}{\alpha}\right)} \\
        \\
        \left(\frac{t-t_{\mathrm{lag}}-t_0}{\tau_W} - \frac{t-t_0}{\tau_W} \right) \left(\frac{t-t_{\mathrm{lag}}-t_0}{\tau_W} + \frac{t-t_0}{\tau_W} \right) &= \ln \left(\frac{-\beta}{\alpha} \right) \\
        \\
        -\frac{t_{\mathrm{lag}}}{\tau_W} \left(\frac{2t-2t_0-t_{\mathrm{lag}}}{\tau_W} \right) = \ln\left(\frac{-\beta}{\alpha} \right) \\
        \\
        \frac{t-t_0-t_{\mathrm{lag}}/2}{\tau_W} = -\ln \left(\frac{-\beta}{\alpha} \right) \frac{\tau_W}{2t_{\mathrm{lag}}} \\
        \\
        t^{\mathrm{max}} = t_0 + \frac{t_{\mathrm{lag}}}{2} - \frac{\tau_W^2}{2t_{\mathrm{lag}}} \ln \left( \frac{-\beta}{\alpha} \right)
    \end{split}
\end{equation}

\end{comment}

\begin{align}
    \alpha \exp\left[-\left(\frac{t-t_0}{\tau_W} \right)^2 + \left( \frac{t-t_{\mathrm{lag}}-t_0}{\tau_W} \right) ^2\right] &= -\beta \notag \\[1em]
    -\left(\frac{t-t_0}{\tau_W} \right)^2 + \left( \frac{t-t_{\mathrm{lag}}-t_0}{\tau_W} \right)^2 &= \ln{\left(\frac{-\beta}{\alpha}\right)} \notag \\[1em]
    \left(\frac{t-t_{\mathrm{lag}}-t_0}{\tau_W} - \frac{t-t_0}{\tau_W} \right) \left(\frac{t-t_{\mathrm{lag}}-t_0}{\tau_W} + \frac{t-t_0}{\tau_W} \right) &= \ln \left(\frac{-\beta}{\alpha} \right) \notag \\[1em]
    -\frac{t_{\mathrm{lag}}}{\tau_W} \left(\frac{2t-2t_0-t_{\mathrm{lag}}}{\tau_W} \right) &= \ln\left(\frac{-\beta}{\alpha} \right) \notag \\[1em]
    \frac{t-t_0-t_{\mathrm{lag}}/2}{\tau_W} &= -\ln \left(\frac{-\beta}{\alpha} \right) \frac{\tau_W}{2t_{\mathrm{lag}}} \notag \\[1em]
    t^{\mathrm{max}} &= t_0 + \frac{t_{\mathrm{lag}}}{2} - \frac{\tau_W^2}{2t_{\mathrm{lag}}} \ln \left( \frac{-\beta}{\alpha} \right) \label{eq:t_max_supplemental}
\end{align}
If ENSO variability returns to its reference value in steady-state (i.e., $\alpha=-\beta$), this turns into $t^{\mathrm{max}}=t_0+t_\mathrm{lag}/2$, which matches our intuitive prediction for a smooth warming trajectory discussed above. When $\alpha \neq -\beta$, there is a correction term whose sign depends on the ratio of these two parameters, corresponding to the contribution of the evolving equilibrium term. 

It is now straightforward to calculate the maximum amplitude:

\begin{equation}
    \Delta \mathrm{M}_{\mathrm{ENSO}}^{\mathrm{max}} = \frac{\Delta T_{\infty}}{2} \left[\left(\alpha+\beta\right) + \alpha \mathrm{erf} \left(\frac{t_\mathrm{lag}}{2\tau_W} - \frac{\tau_W}{2t_{\mathrm{lag}}} \ln \left(\frac{-\beta}{\alpha} \right) \right) + \beta \mathrm{erf} \left(-\frac{t_\mathrm{lag}}{2\tau_W} - \frac{\tau_W}{2t_{\mathrm{lag}}} \ln \left(\frac{-\beta}{\alpha} \right) \right) \right]
    \label{eq:ENSOAmplitudeAnalytic}
\end{equation}
which is a prediction of maximum ENSO amplitude in terms of the total amount of warming, a warming timescale, and the three lag-linear parameters. We can once again gain intuition about what this expression means by considering the case when $\alpha=-\beta$. The corrections to the arguments of the error functions disappear, as does the initial $\alpha+\beta$ representing the long-term warming effect (because it is multiplied by $\Delta T_{\infty}$). This leaves:

\begin{equation}
    \Delta \mathrm{M}_{\mathrm{tr}}^{\mathrm{max}} = -\beta \Delta T_{\infty} \mathrm{erf} \left(\frac{t_{\mathrm{lag}}}{2\tau_W} \right)
\end{equation}
which, as expected, is proportional to a normalized warming amplitude and depends on the ratio of timescales $t_{\mathrm{lag}}/\tau_W$.

We can also write an expression for peak ENSO amplitude using non-dimensional parameters. The first parameter appears in nearly every expression; $\lambda\equiv t_\mathrm{lag}/\tau_W$ tells us over how many subtropical cell timescales the warming occurs. This strongly impacts the magnitude of the ENSO peak; if the subsurface warming cannot keep up with the surface warming then El Ni\~no temperatures will increase faster than La Ni\~na temperatures. The non-dimensionalized warming timescale can affect the timing of the peak as well, but only does so significantly if there is a strong steady-state response (i.e., $-\beta/\alpha$ is far from one) or $\lambda$ is small.

The second non-dimensional parameter is $\rho=-\beta/\alpha$, and appears when we factor out a non-dimensional temperature change ($\theta=\alpha \Delta T_{\infty}$) from the right side of Eq. \ref{eq:ENSOAmplitudeAnalytic}:

\begin{equation}
    \Delta \mathrm{M}_{\mathrm{ENSO}}^{\mathrm{max}} = \frac{\theta}{2} \left[1-\rho + \mathrm{erf} \left( \frac{\lambda}{2} - \frac{\ln \rho}{2\lambda}  \right) +\rho \mathrm{erf} \left(\frac{\lambda}{2} + \frac{\ln{\rho}}{2\lambda} \right) \right]
\end{equation}
We can also non-dimensionalize the time of the peak:

\begin{equation}
    \frac{t^{\mathrm{max}}-t_0}{t_\mathrm{lag}} = \frac{1}{2} - \frac{ \ln{\rho}}{2 \lambda^2}.
\end{equation}

These predictions are shown in Extended Data Fig. 6. ENSO variability is always proportional to the non-dimensionalized temperature change ($\theta$), meaning more warming leads to a larger peak. Additionally, larger values of $\lambda$ (shorter warming timescales) correspond to larger peak variability, and if $\rho=1$ this has no impact on peak timing. If $\rho<1$, peaks are shifted later and larger, while if $\rho>1$, peaks are shifted earlier and smaller.

\subsection{Adjusting for Abrupt Warming}

When an abrupt change in radiative forcing occurs, it is generally thought that global mean surface temperatures follow a multi-exponential curve corresponding to heat uptake on 1. a fast surface timescale $\tau_W$ and 2. a slow deep-ocean adjustment timescale $\tau_s$ \citep{geoffroy_transient_2013}. We will assume a double-exponential:

\begin{equation}
    \Delta \overline{\mathrm{GMST}} \;=\; \begin{cases}
0, & t \le 0,\\
(\Delta T_{\infty}-A_s)(1-e^{-t/\tau_W}) + A_s(1-e^{-t/\tau_s}), & t > 0,
\end{cases}
\label{eq:twoexp}
\end{equation}
where $A_s$ is the amplitude of the slow warming. The derivative of temperature is a finite positive value at $t=0^+$, and smoothly decreases after that. Therefore, the most warming that occurs in a period of length $t_\mathrm{lag}$ is immediately after warming begins, leading to the straightforward conclusion that $t^{\mathrm{max}}_{\mathrm{tr}}=t_\mathrm{lag}$, and other than in some unrealistic cases where the equilibrium term dominates, $t^{\mathrm{max}}=t_\mathrm{lag}$. The maximum amplitude is therefore:

\begin{equation}
    \Delta \mathrm{M}^{\mathrm{max}}_{\mathrm{ENSO}} = (\Delta T_{\infty} -A_s) (1-\exp \left(-t_{\mathrm{lag}}/\tau_W \right)) + A_s(1-\exp(-t_\mathrm{lag}/\tau_s))
\end{equation}

While this expression is not as intuitive or useful as that of the smooth warming case, it is clear that longer subtropical timescales or shorter warming timescales enhance ENSO peaks, as does the total warming amplitude. It is also a useful testing ground for what the lag-linear model predicts in a different temperature trajectory, and fits simulations well (Fig. 5b.ii). 

\clearpage

\section*{Supplemental Figures}{\label{sec:Supplemental}}

\begin{figure}[h]
\centering
\includegraphics[width=0.9\textwidth]{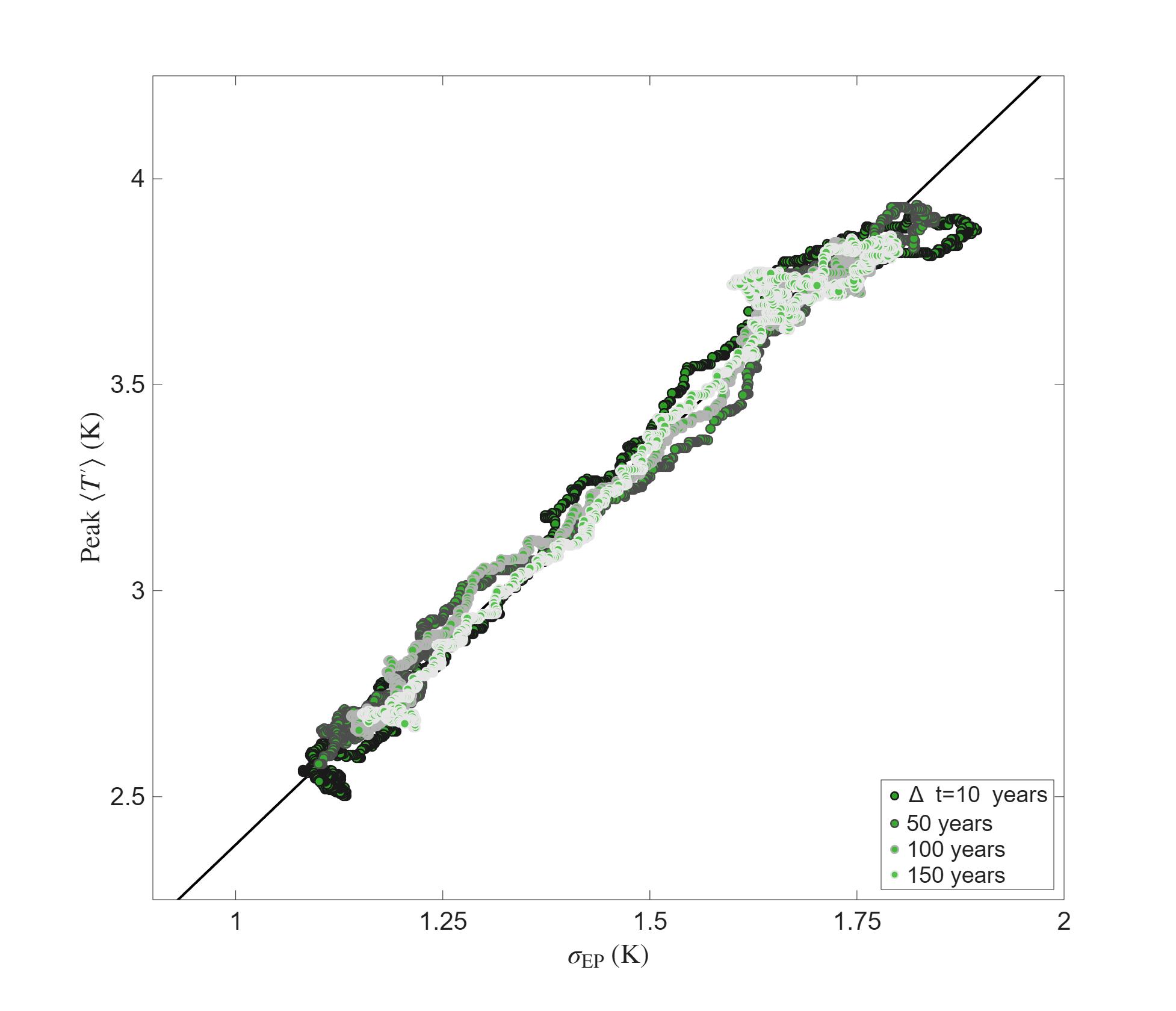}
\caption{The relationship between East Pacific standard deviation (x) and peak volume average anomalous East Pacific temperature (y) over time. The ensemble mean of each warming timescale are shown as different colors.}\label{Fig:SupplementalPeakVsStandardDeviation}
\end{figure}

\begin{figure}[h]
\centering
\includegraphics[width=0.9\textwidth]{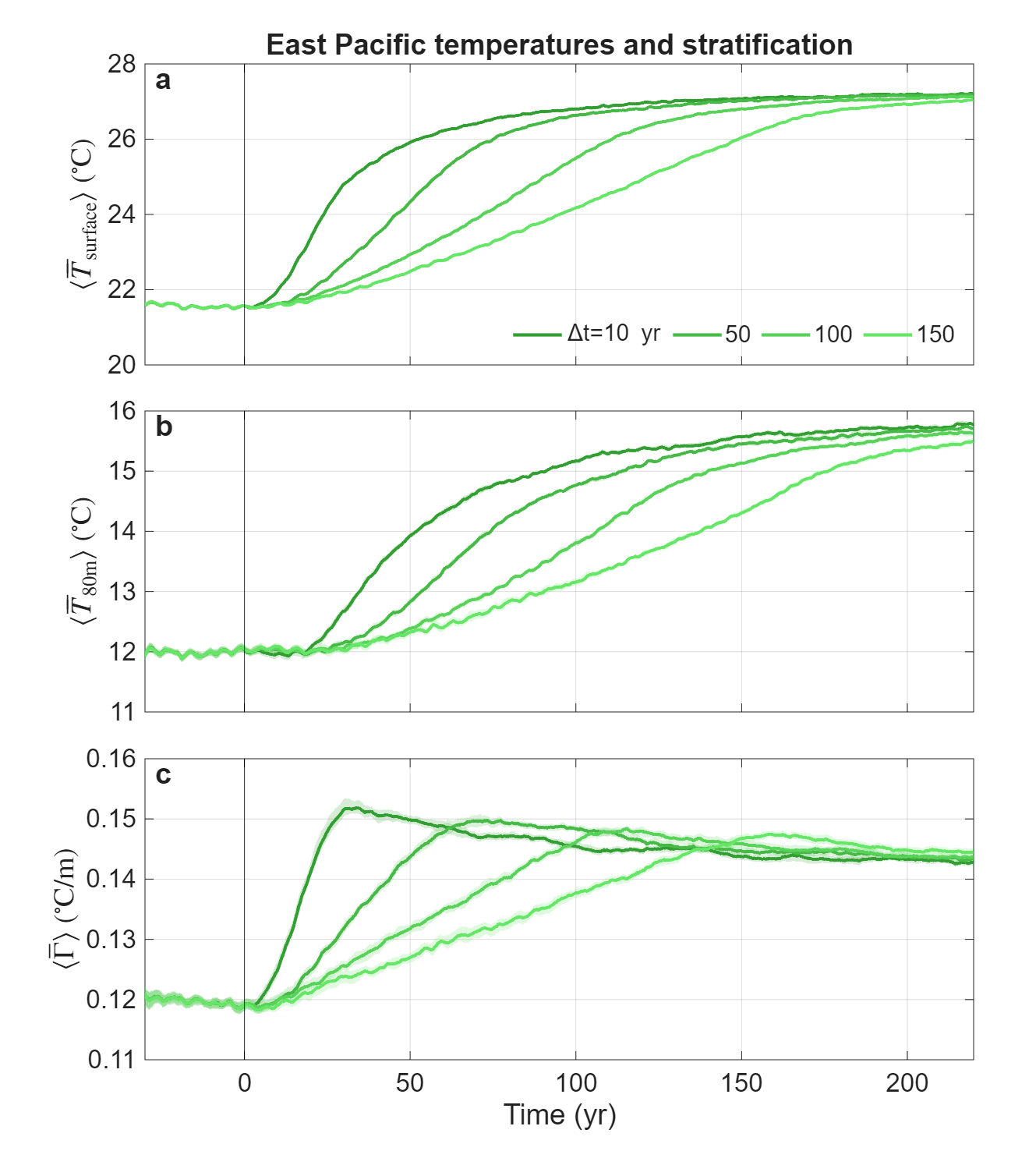}
\caption{Panel a shows the East Pacific surface temperature, panel b the subsurface temperature, and panel c the stratification over time. The ensemble mean of each warming timescale are shown as different colors.}\label{Fig:SupplementalSurfaceAndDepth}
\end{figure}

\begin{figure}[h!]
\centering
\includegraphics[width=0.8\textwidth]{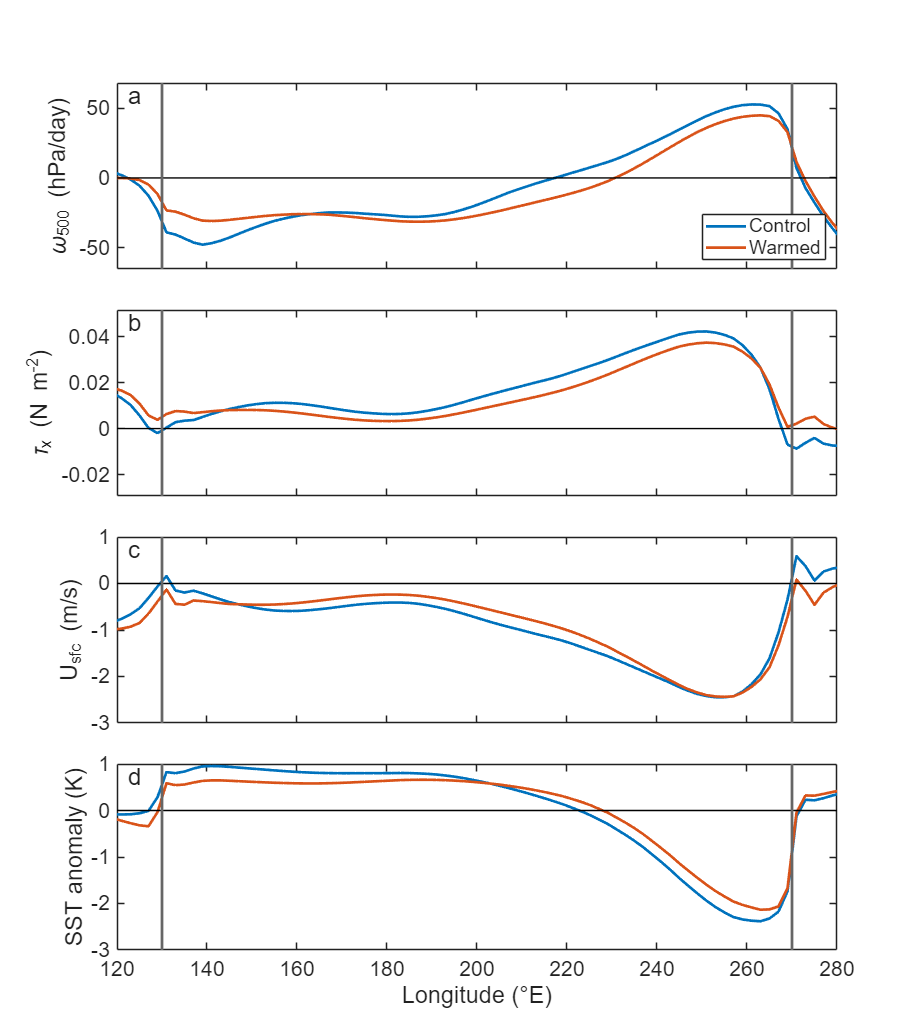}
\caption{Changes to the Walker circulation with warming. Panel a shows free tropospheric vertical velocity at $\sim$500 hPa, panel b shows surface wind stress, panel c shows surface wind, and panel d shows the SST difference from the zonal mean. Wind speeds are averaged within 5$^\circ$ of the equator and compared between the reference simulation (blue) and the $\Delta t=10$yr warming simulations (red). The ascending motion ($\sim$130-150$^\circ$E), descending motion ($\sim$260$^\circ$E), and trade winds ($\sim$160-230$^\circ$E) are all weaker in a warmer climate, leading to a shallower temperature gradient.}\label{Fig:WalkerWeakening}
\end{figure}

\begin{figure}[h]
\centering
\includegraphics[width=\textwidth]{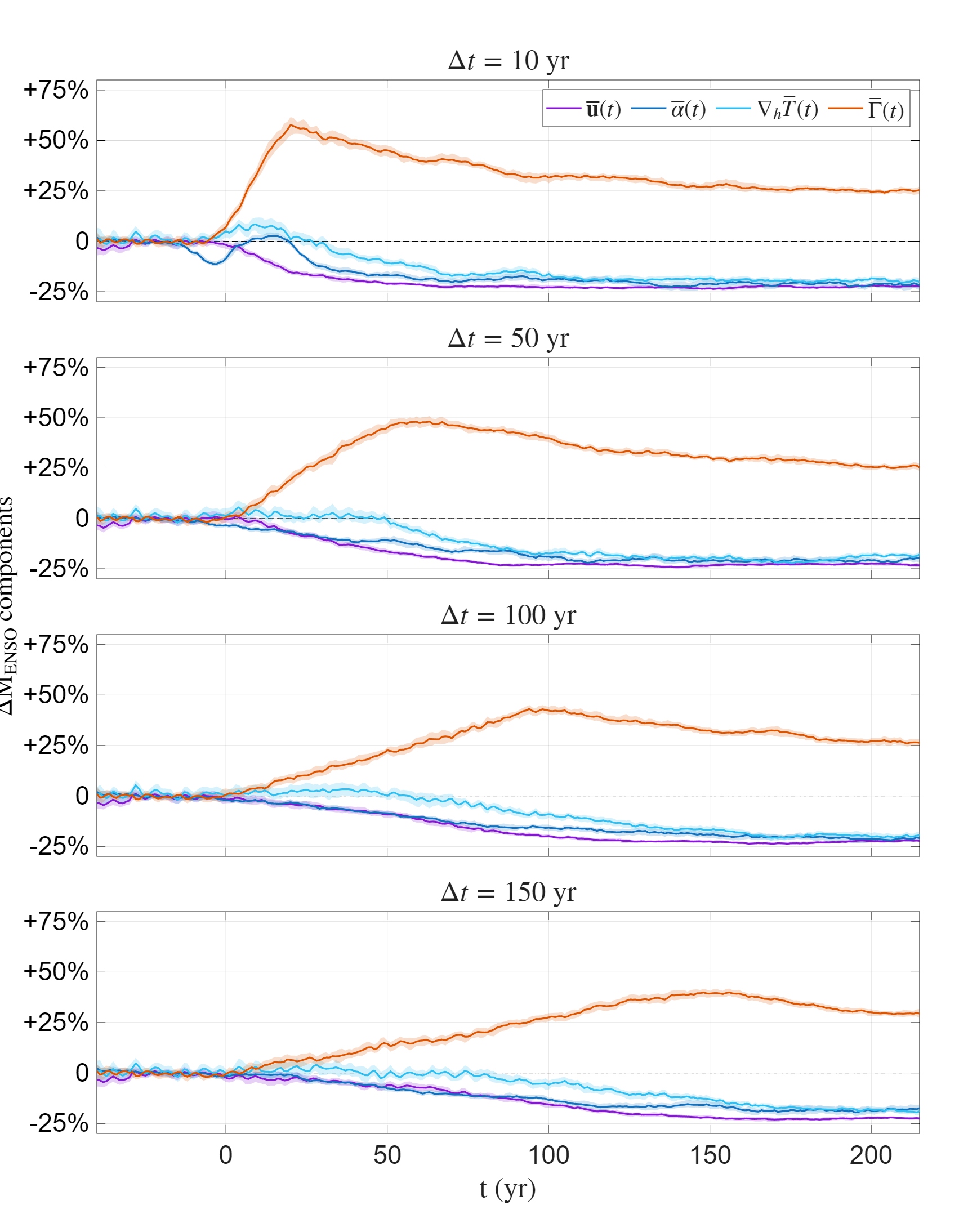}
\caption{Same as Fig. 3d, but for each warming timescale.}\label{Fig:SupplementalBudgetComponentsEachdt}
\end{figure}

\begin{figure}[h]
\centering
\includegraphics[width=\textwidth]{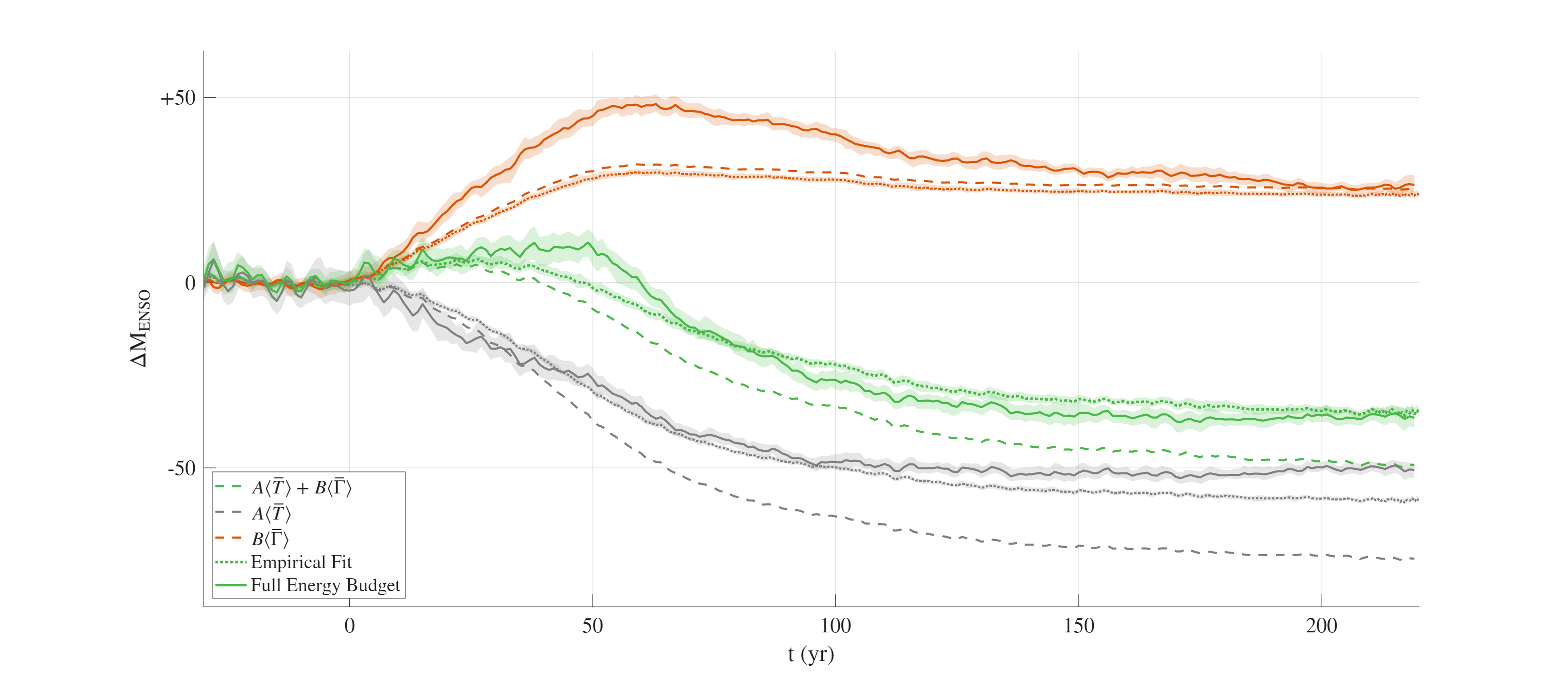}
\caption{ENSO magnitude over time, comparing the energy budget approximations given by Eq. 6 (dashed), the empirically fit linear model (dotted), and the full energy budget (solid). For each of the three, the result of changing stratification (red), mean temperature (black), or the full time evolution (green) are shown.}\label{Fig:SupplementalEnergyBudgetPredictions}
\end{figure}

\begin{figure}[h]
\centering
\includegraphics[width=\textwidth]{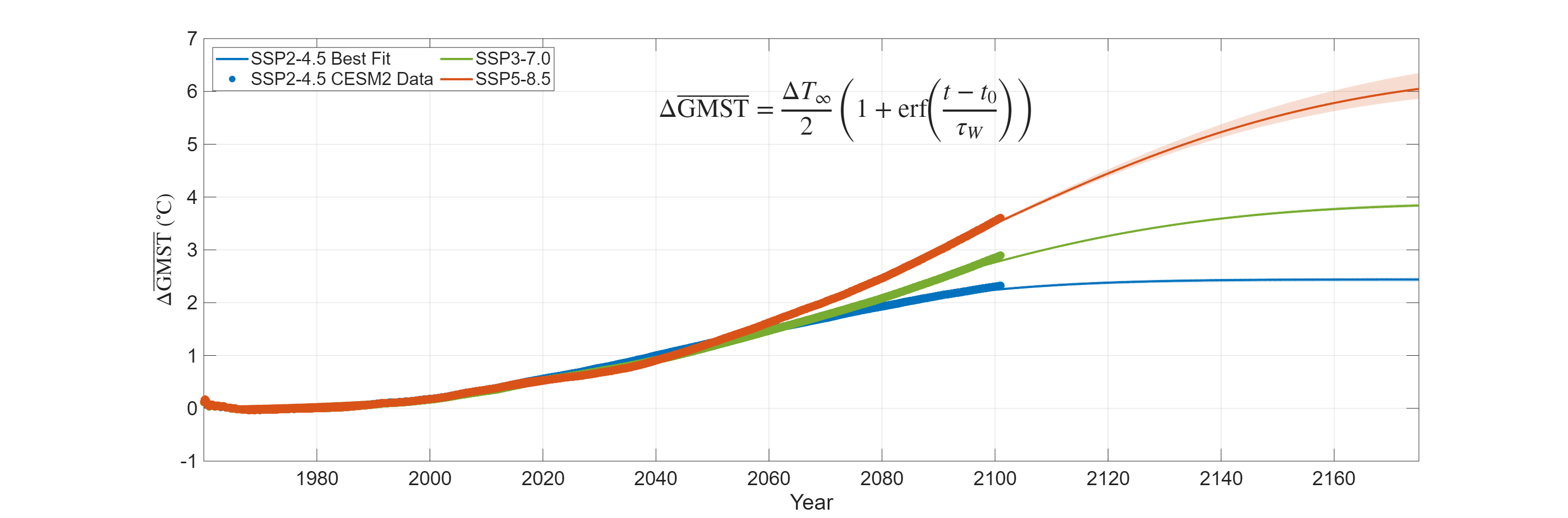}
\caption{Sample idealized temperature trajectories, capturing the evolution of GMST with a few parameters, and how they compare to various scenarios simulated by CESM2.}\label{Fig:SupplementalGMSTTrajectories}
\end{figure}

\begin{figure}[h]
\centering
\includegraphics[width=\textwidth]{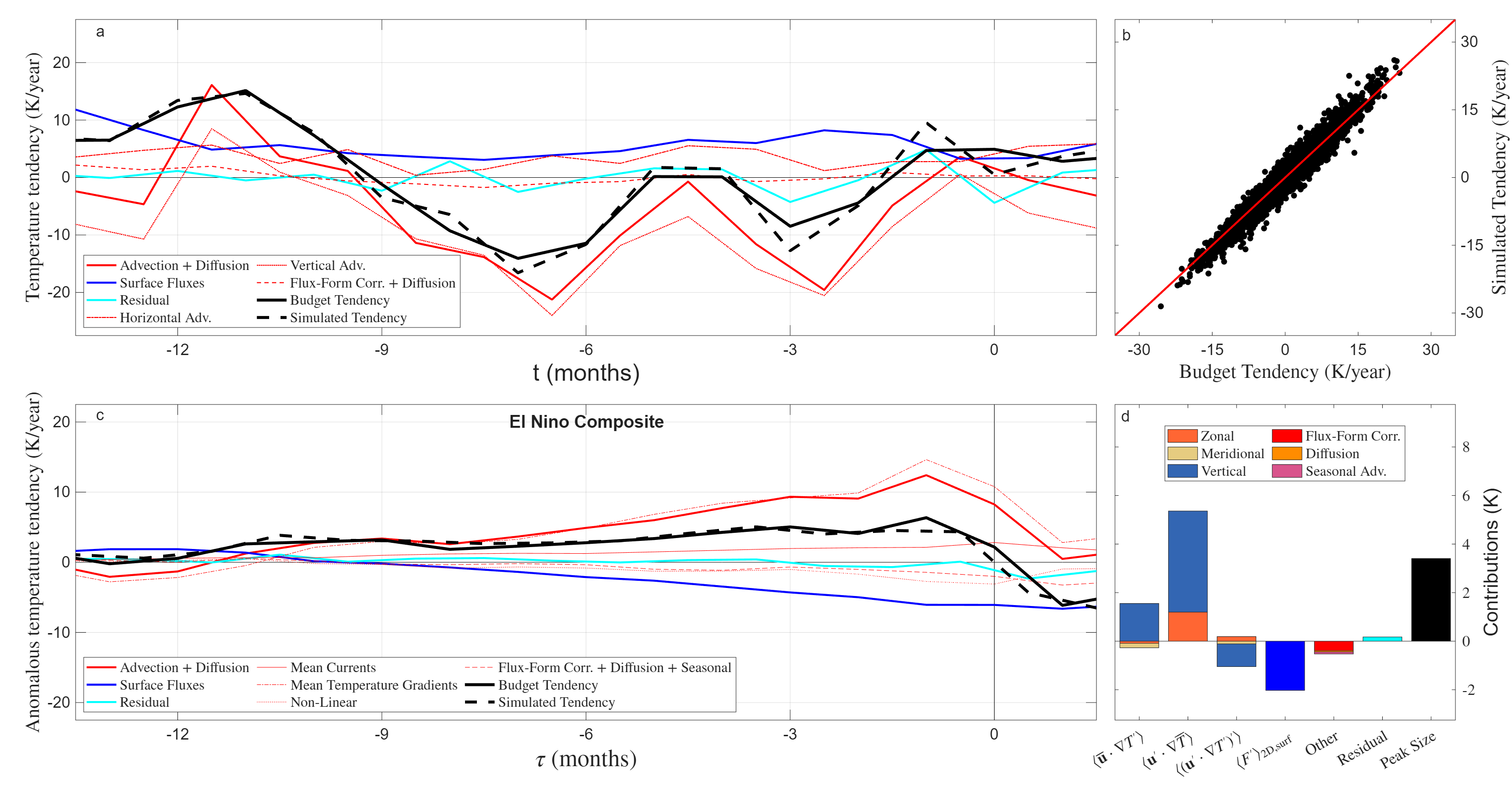}
\caption{Temperature budget calculations for the East Pacific mixed layer. Panel a shows a sample time series of simulated temperature tendency (dashed black line), its advective (red) and surface flux (blue) contributions, and the resulting predicted budget tendency (solid black). The residual (cyan) and advection sub-components (thin red lines; horizontal, vertical, and seasonal cycle and flux form correction) are also shown. Panel b shows the degree to which the budget tendency matches simulated temperature changes at all times in the control simulation. Panel c shows a similar budget for an El Ni\~no composite as a function of time relative to the peak $\tau$. Each quantity is averaged over 25 events from the control simulation, chosen as those with the maximum values of $\langle T' \rangle$ divided by the moving standard deviation separated by at least two years. Panel d shows the integral of each term up to $\tau=0$, i.e., their contribution to the total El Ni\~no peak.}\label{Fig:SupplementalTBudget}
\end{figure}

\begin{figure}[h]
\centering
\includegraphics[width=0.9\textwidth]{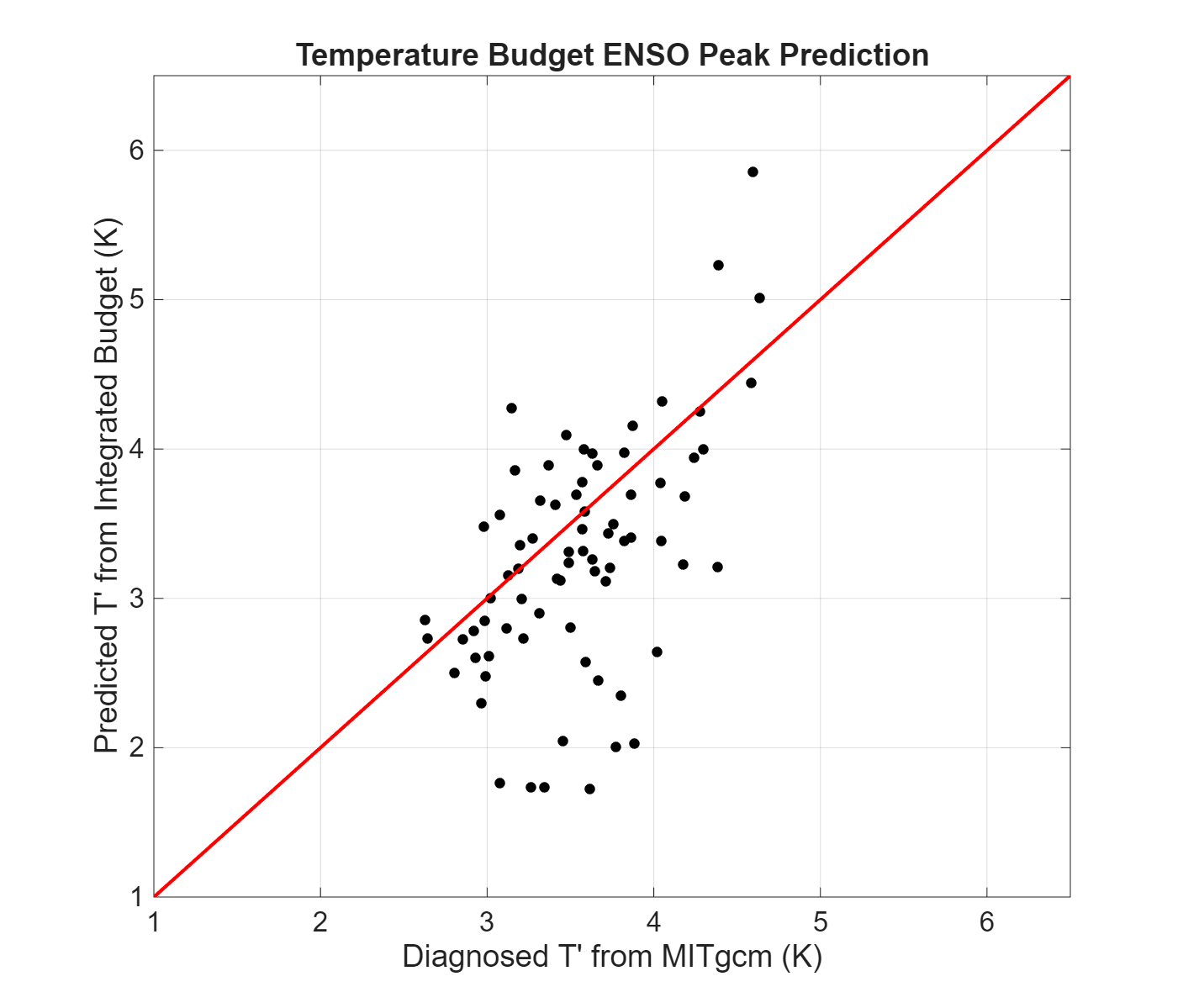}
\caption{The maximum $T'$ achieved by simulated El Ni\~no peaks compared to those predicted from the budget discussed and the identity line (x=y, red) as a basis of comparison. 25 El Ni\~no events from the control simulation are shown and two El Ni\~no events from the pre-warming period of each other simulation. For this figure $\tau_0$ is defined as when $\langle T' \rangle$ crosses 1 K.}\label{Fig:SupplementalSimulatedVsPredictedPeaks}
\end{figure}

\begin{figure}[h]
\centering
\includegraphics[width=0.9\textwidth]{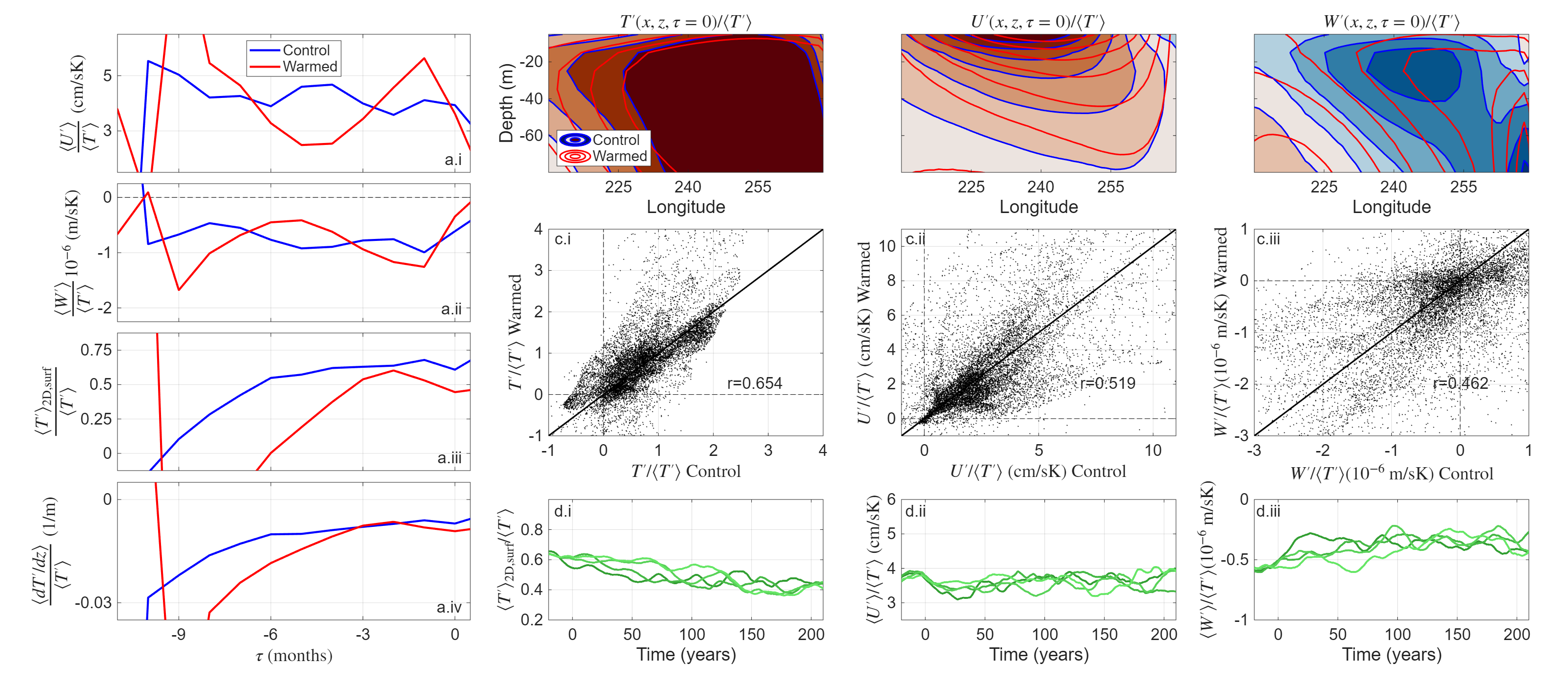}
\caption{How the structure of ENSO events change under warming. Panel a shows the volume averaged anomalous zonal (a.i) and vertical currents (a.ii), surface temperature (a.iii), and stratification (a.iv) as a function of $\tau$, all normalized by the anomalous temperature and shown for the quasi-steady-state control and warmed climates. Panel b shows longitude-depth slices at $\tau=0$ of anomalous temperature (b.i), zonal current (b.ii), and vertical current (b.iii), also normalized and displayed for the control and warmed climates. Panel c shows the correlation between the control and warmed normalized anomalies for the same three quantities across all values of $\tau$, $x$, $y$,and $z$ in the relevant region and after $\tau_0$. Lastly, panel d shows how relevant anomalous quantities change over time, where the value at each time is calculated from the composite of the 20 ENSO events closest to that point in time across all simulations with the same warming timescale. As with the figures in the main text, the color corresponds to the warming timescale. }\label{Fig:ENSOStructureWarming}
\end{figure}

\begin{figure}[h]
\centering
\includegraphics[width=0.9\textwidth]{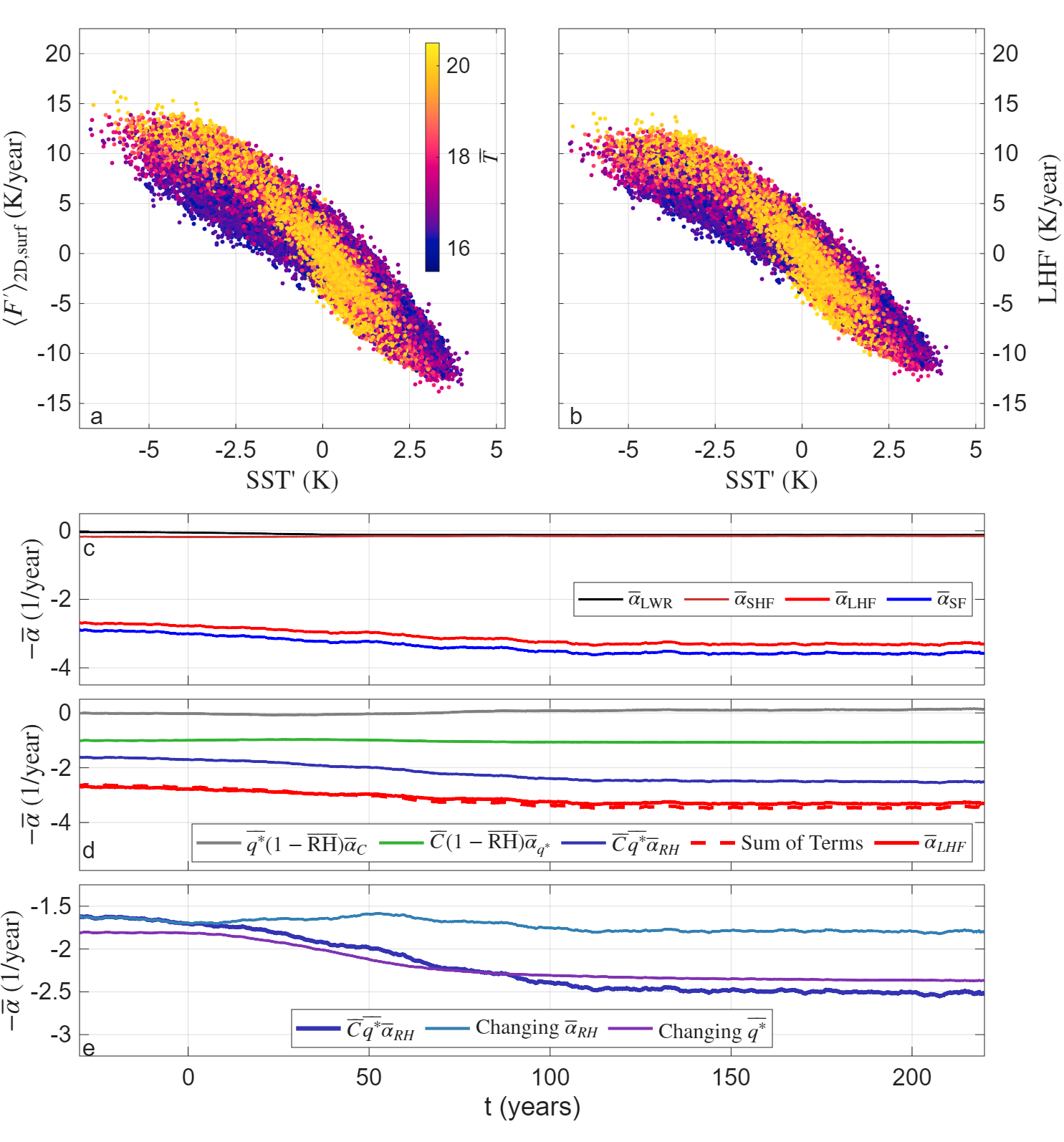}
\caption{Dependence of surface flux damping and its components on warming. Panel a shows the sum of anomalous radiation and turbulent fluxes (i.e., $\langle{F}'\rangle_{\text{2D,surf}}$) as a function of anomalous surface temperature across all MITgcm simulations, with color representing mean temperature. Panel b shows the same, but the temperature lost through only anomalous latent heat flux (LHF) rather than all sub-grid-scale processes. Panel c shows how the relationship between the quantities shown in a and b change over time in the $\Delta t=50$ year simulations: $\bar{\alpha}$ is diagnosed as the slope between an anomalous energy source (or sink) and temperature over a 20 year period for positive values of SST$'$, and the contributions from LHF, sensible heat (SHF), and radiation (LWR), as well as the total $\bar{\alpha}_{\mathrm{SF}}$, are shown. Panel d shows the linear components of $\bar{\alpha}_{\mathrm{LHF}}$ and their sum, while panel e further decomposes the most relevant term into changes due to saturation specific humidity and anomalous relative humidity.}\label{Fig:SupplementalSurfaceFluxes}
\end{figure}

\begin{figure}[h]
\centering
\includegraphics[width=0.9\textwidth]{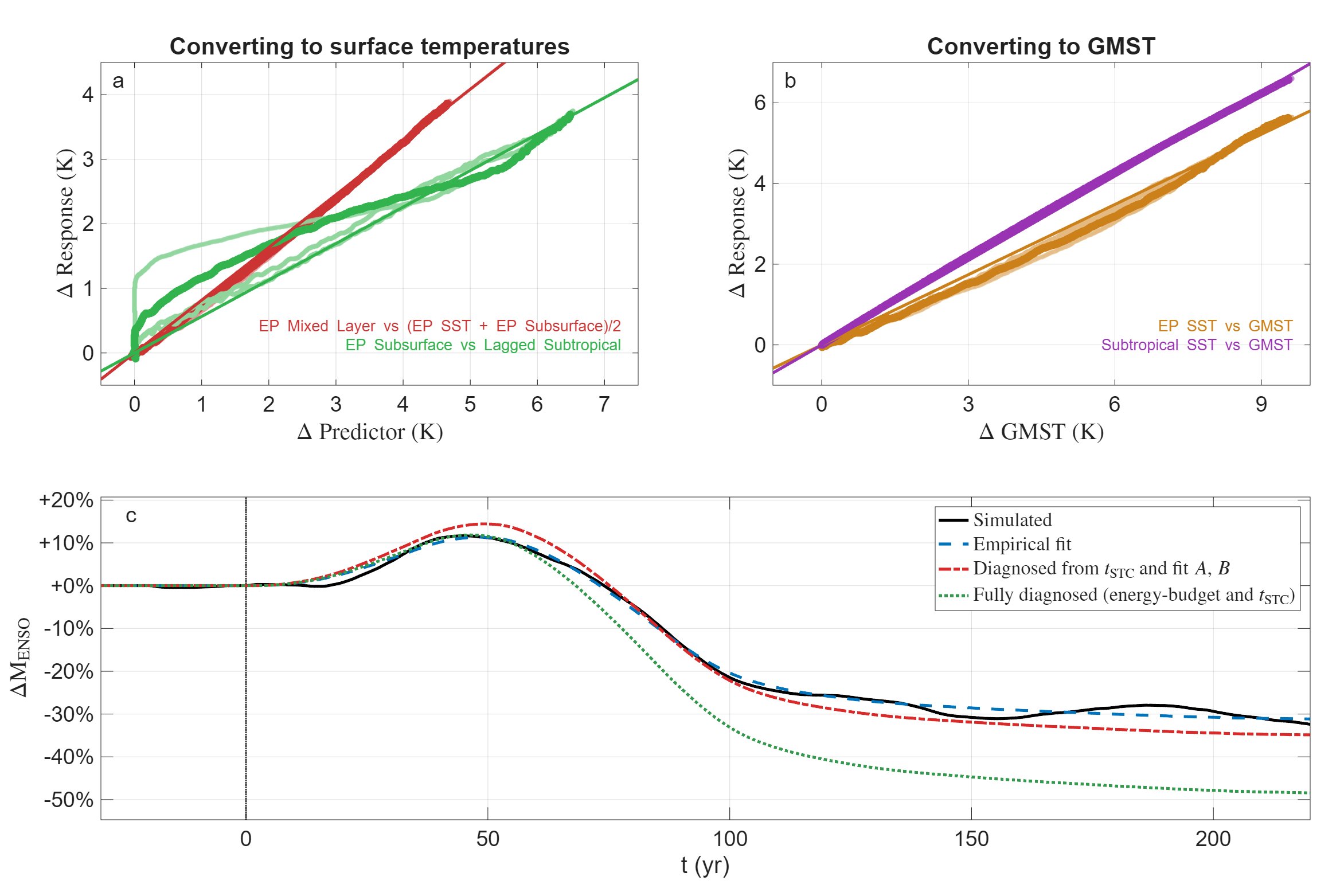}
\caption{The relevant linear relationships for connecting our two linear models and the resulting predictions of ENSO over time. Panel a shows the connections between the first linear model arguments and surface temperatures, while panel b connects surface temperatures to GMST. In each panel, all four $\Delta t$ ensembles and the line of best fit are shown, with the $\Delta t=$50 yr simulation shown more brightly. Panel c uses the diagnosed slopes and the subtropical cell timescale to predict ENSO variability, starting from either the fit $A$ and $B$ values or the diagnosed ones. These are compared to the empirical fit and simulated ENSO variability.}\label{Fig:GMSTProportionality}
\end{figure}

\clearpage

%\include{flowchart_v3}

\begin{comment}
\begin{figure*}[tbp]
  \centering
  \noindent\makebox[1.05\textwidth][c]{\input{flowchart_v3}}
  \caption{...}
  \label{fig:flowchart}
\end{figure*}
\end{comment}

\bibliography{sn-bibliography}